\documentclass[aps,prx,notitlepage,amsfonts,citeautoscript,superscriptaddress,showpacs,twocolumn,print]{revtex4-1}

\usepackage{natbib}
\usepackage{booktabs}
\usepackage{listings}
\usepackage{color}
\definecolor{dkgreen}{rgb}{0,0.6,0}
\definecolor{gray}{rgb}{0.5,0.5,0.5}
\definecolor{mauve}{rgb}{0.58,0,0.82}

\usepackage{amsmath,amssymb,mathrsfs}
\usepackage{latexsym}
\usepackage{graphicx} 
\usepackage{grffile}
\usepackage[normalem]{ulem}
\usepackage{epstopdf}
\usepackage{graphicx,epstopdf,color}
\usepackage{amsfonts}
\usepackage{amsmath,amssymb,mathrsfs}
\usepackage{hyperref}
\usepackage{mhchem}   
\usepackage{siunitx}  

\usepackage[dvipsnames]{xcolor}
\usepackage{url}
\usepackage{soul}
\usepackage{cancel}
\usepackage{appendix}
\usepackage{cleveref}
\usepackage{mhchem}

\crefname{equation}{Eq.}{Eqs.}
\Crefname{equation}{Equation}{Equations}
\crefname{figure}{Fig.}{Figs.}
\Crefname{figure}{Figure}{Figures}
\crefname{section}{Sec.}{Secs.}
\crefname{subsection}{Subsec.}{Subsecs.}
\Crefname{section}{Section}{Sections}
\crefname{appendix}{Appendix}{Apps.}
\crefname{paragraph}{Sec.}{Secs.}
\crefname{table}{Table}{Tables}

\usepackage{bm}
\newcommand{\textalert}[1]{}

\newcommand{\gens}{g_{\text {ens }}} 
\newcommand{\bargens}{\mathfrak{g}} 
\newcommand{\ks}{\kappa_{\mathrm{s}}}

\newcommand{\Ein}{\varepsilon_{\text {in}}}
\newcommand{\Eout}{\varepsilon_{\text{out}}}

\newcommand{\Eein}{\mathcal{E}_{\text {in}}}

\newcommand{\half}{\frac{1}{2}}

\def\ie{i.e.\ }

\usepackage{xr}

\graphicspath{{nb_fig/}}

\usepackage[caption=false]{subfig}
\usepackage{epstopdf}
\usepackage{xcolor}

\makeatletter
\def\@fnsymbol#1{\ensuremath{\ifcase#1\or * \or \mathsection\or \mathparagraph\or \|\or **\or \else\@ctrerr\fi}}
\makeatother

\begin{document}
\title{Optimal absorption and emission of itinerant fields into a spin ensemble memory}

\author{Linda Greggio}
\thanks{linda.greggio@inria.fr}
\affiliation{Laboratoire de Physique de l’Ecole Normale Supérieure, Mines Paris, Inria,
CNRS, ENS-PSL, Sorbonne Université, PSL Research University, Paris, France}
\author{Tristan Lorriaux}
\affiliation{Laboratoire de Physique de l’Ecole Normale Supérieure de Lyon, France}
\author{Alexandru Petrescu}
\affiliation{Laboratoire de Physique de l’Ecole Normale Supérieure, Mines Paris, Inria,
CNRS, ENS-PSL, Sorbonne Université, PSL Research University, Paris, France}

\author{Mazyar Mirrahimi}
\affiliation{Laboratoire de Physique de l’Ecole Normale Supérieure, Mines Paris, Inria,
CNRS, ENS-PSL, Sorbonne Université, PSL Research University, Paris, France}
\author{Audrey Bienfait}
\thanks{audrey.bienfait@ens-lyon.fr}
\affiliation{Laboratoire de Physique de l’Ecole Normale Supérieure de Lyon, France}
\date{\today}
\begin{abstract}
Quantum memories integrated in a modular quantum processing architecture can rationalize the resources required for quantum computation. This work focuses on spin-based quantum memories, where itinerant electromagnetic fields are stored in large ensembles of effective two-level systems, such as atomic or solid-state spin ensembles, embedded in a cavity. Using a mean-field framework, we model the ensemble as an effective spin communication channel and describe both absorption and emission processes using a cascaded quantum model. We derive optimal time-dependent modulations of the cavity linewidth that maximize storage and retrieval efficiency for fast incoming pulses. Our analysis yields an upper bound on efficiency, which can be met in the narrow bandwidth regime. It also shows the existence of a critical bandwidth above which the efficiency severely decreases. Numerical simulations are presented in the context of microwave-frequency quantum memories interfaced with superconducting quantum processors, highlighting the protocol's relevance for modular quantum architectures.
\end{abstract}
\maketitle

\section{Introduction}
Quantum memories - devices capable of faithfully capturing, storing and retrieving one or more quantum states - are fundamental components in a range of quantum technologies \cite{heshami2016}. Their ability to preserve quantum information can enhance long-distance communication~\cite{kimble2008,duan2001,bhaskar2020} through quantum repeaters~\cite{sangouard2011}, increase the sensitivity of quantum sensors~\cite{zaiser2016,ding2020} and reduce the resource requirements for quantum computation~\cite{thaker2006,gouzien2021}. Various physical systems can serve as quantum memories, including atomic ensembles~\cite{lvovsky2009}, solid-state defects~\cite{tittel2010,clausen2011}, molecular gases~\cite{rabl2006}, single-mode oscillators~\cite{reagor2016,bozkurt2024}, atoms and ions~\cite{sangouard2009,reiserer2015}. These systems rely on interactions such as microwave~\cite{reagor2016}, mechanical~\cite{bozkurt2024}, electronic~\cite{lvovsky2009}, magnetic~\cite{bradley2019}, or vibrational coupling~\cite{tittel2010} to store quantum information. This article focuses on one particular approach: storing itinerant electromagnetic fields using the inhomogeneous broadening of an ensemble of effective two-level systems (spins) addressed with a cavity, such as atoms or solid-state defects, and retrieving them using echo-based protocols~\cite{channeliere2018}.
Proposals~\cite{AfzeliusOptics, Afzelius_2013,julsgaard} for quantum memories based on inhomogeneously-broadened spins in a cavity often emphasize operating in a “perfectly matched” regime, where to achieve efficient absorption, the losses induced by the ensemble on the cavity precisely balance the decay rate of the cavity due to its coupling to the input line. This efficiency sweet-spot was derived considering “slow” incoming wavepackets, whose bandwidth was considerably smaller than the spin ensemble linewidth, the cavity linewidth, and the cavity to spin ensemble coupling. However, practical quantum memories should be able to exchange information with quantum processors~\cite{gouzien2021} with a high swap rate. For absorption-based quantum memory, this requires realizing quantum state transfers (QST) between the memory and the processor qubits with short-duration wavepackets. Between qubits, efficient QSTs require to dynamically control the coupling strength between the communication channel and the emitter node, as well as the receiver node~\cite{Jahne2007,korotkov2011, chatterjee2022}. 
Here, similarly to~\cite{Bernad2025}, we show that the same strategy can apply for quantum memories operating with fast incoming pulses. We cast the spins-cavity system during the absorption and emission stages of the memory to a quantum cascaded system~\cite{Gardiner93}. This framework allows us to determine the maximum achievable storage efficiency as a function of the memory's physical parameters and the bandwidth of the incoming quantum signal. Furthermore, we derive the optimal time-dependent modulation of the cavity linewidth that maximizes storage and retrieval efficiency into the spin ensemble for an incoming pulse of arbitrary duration.

The spin ensemble is characterized by an inhomogeneous linewidth $\Gamma$,  determined by the frequency distribution of the spins.  This linewidth sets both the spectral bandwidth of the memory and the upper bound on its operational speed ~\cite{channeliere2018}. Upon absorption, a quantum state becomes rapidly scattered across the internal degrees of freedom of the ensemble.  Retrieval protocols are typically based on spin-echo strategies~\cite{tittel2010, damon2011, julsgaard,O'SullivanPRX2022}, in which an initial magnetization created in the ensemble is refocused by applying pulse sequences that counteract dephasing from inhomogeneous broadening. Successful retrieval is limited by the ensemble's coherence time $T_2$. Depending on the spin and its dominant decoherence mechanisms, a variety of refocusing pulse sequences — ranging from simple Hahn echoes to more elaborate multi-pulse schemes — have been developed to extend coherence and thus storage time~\cite{grezes2015}. The ratio of the temporal duration of the stored signal to the total coherence time offers one metric for a key figure of merit: the quantum memory capacity, which quantifies how many distinct temporal modes can be reliably stored. 

In contrast to these refocusing pulses, which impose their own constraints on the physical characteristics of the memory and its operation, optimizing the absorption and emission efficiency primarily depends on achieving sufficient coupling between the spin ensemble and the cavity. In such hybrid systems, the loss rate induced by the spins on the cavity is given by $\ks =4 \gens^2/\Gamma$~\cite{staudt2012}, where $\gens$ is the collective coupling strength between the cavity and the ensemble. Efficient absorption of slow or continuous signals — those with bandwidths smaller than the inhomogeneous broadening — can be achieved by matching $\ks$ to the cavity’s decay rate $\kappa$, assuming this rate to be solely governed by the cavity coupling to the input channel, with no additional intrinsic losses.
This has been shown in earlier mean-field analyses \cite{Afzelius_2013,julsgaard}.  In this article, we extend this line of work to explore the storage and retrieval efficiency of fast itinerant pulses with two primary objectives. First, we aim to increase the storage capacity of the memory. Second, we seek to better characterize the absorption and emission timescales of the memory to compare them to the timescale of typical processing quantum nodes, in the context of modular architectures incorporating quantum memories. Finally, we also include the intrinsic cavity loss in this analysis, which turns out to be an important limiting factor for the efficiency of the protocol.

As shown in~\cite{Bernad2025} for the absorption step, efficient storage of incoming pulses requires dynamic modulation of the cavity linewidth, mirroring strategies used in itinerant QST for optimal efficiency. 
We remain within the mean-field theoretical framework developed in earlier works \cite{Afzelius_2013,julsgaard}, and we leverage this description to model the ensemble as an effective spin communication channel. The channel interacts with a fictitious spin bosonic field with coupling rate $\Gamma$, which in turn couples to the cavity with strength $\gens$. This model reduces the absorption process to a system of two coupled bosonic modes with one mode able to decay into a spin communication channel. In particular, it allows us to formulate and solve analytically an optimization problem to maximize absorption efficiency. 
Compared to other approaches~\cite{Bernad2025}, the emission step naturally emerges as a quantum cascade of the absorption step, with the emitted field into the spin channel driving a future instance of the same two-mode system. This enables a similar optimization of the emission process.
 
The article is organized as follows: in a first section, we briefly recall the physics of the itinerant absorption-refocus-retrieve protocol, and introduce the physical system and its interaction structure. In a second part, we map the evolution of this system during the protocol into a quantum cascaded model. From this model, we derive the protocol efficiency in a steady-state operation regime. In \cref{sec:Optimization}, we demonstrate that these steady-state efficiencies serve as upper bounds for the more realistic case of finite-length wavepackets, and derive the optimal modulation functions needed to maximize efficiency for a given input waveform. Finally, we perform numerical simulations to compute these optimal modulation profiles and analyze the corresponding decrease in efficiency as a function of the incoming signal bandwidth and the cavity’s intrinsic loss. We do so in the particular context of quantum memories implemented at microwave frequencies, and possibly coupled to superconducting quantum processors.

\section{Physical model and storage protocol}
\label{sec:Physical model}

\begin{figure}
    \centering
    \includegraphics[width=.9\linewidth]{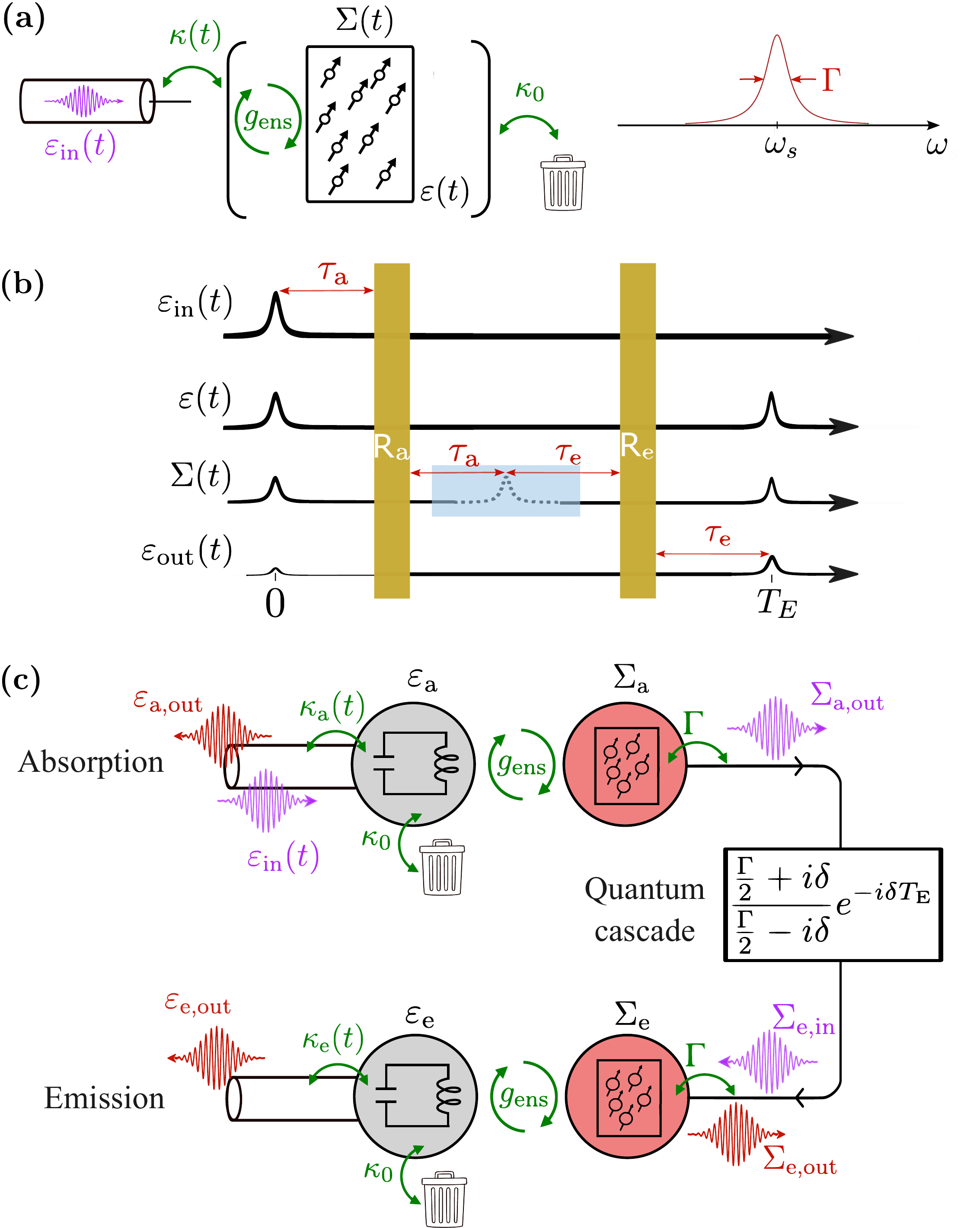}
    \caption{\textbf{(a)} Spin ensemble of central frequency $\omega_\text{s}$ and of inhomogeneous broadening $\Gamma$ coupled with strength $\gens$ to a cavity driven by the input field $\Ein(t).$  The cavity is subject to intrinsic losses $\kappa_0$ and interacts with the transmission line through a tunable coupling $\kappa(t).$ The intra-cavity field $\varepsilon(t)$ and the field in the spin ensemble $\Sigma(t)$ describe the dynamics of the stored field.
    \textbf{(b)} Memory protocol: once the incoming field is absorbed, the spins start to dephase. Two refocusing pulses are used to rephase the spins, and an echo is emitted at time $T_{\text{E}}.$ 
    To avoid the emission of the first noisy echo at $2 \tau_{\text{a}}$, the cavity is detuned between the two pulses. $\Eout(t)$ represents the retrieved field.
    \textbf{(c)} Quantum cascaded formalism: the memory system is represented as two cascaded copies of a two-mode bosonic system. In the absorption copy the incoming field $\Ein$ is stored in the spin ensemble, while in the emission copy the spin ensemble is driven by a feedback term $\Sigma_{\text{e,in}}$, representing a filtered version of the absorption output field $\Sigma_{\text{a,out}}$, with filter function specified in the box.}
    \label{fig:1}
\end{figure}

An ensemble of $N$ spins is embedded in a cavity with intrinsic loss rate $\kappa_0$, see \cref{fig:1}\textbf{(a)}. The cavity field is described by its annihilation $\hat{\varepsilon}$ and creation $\hat{\varepsilon}^\dagger$ operators, and is coupled to a measurement waveguide at a tunable coupling rate $\kappa(t)$.  Any spin $j$ of the ensemble, of Larmor frequency $\omega_j$, interacts with the cavity at a strength $g_j$ with an interaction Hamiltonian $\hat{H}_j/\hbar = g_j (\hat{\varepsilon} \hat{\sigma}^j_+ + \hat{\varepsilon}^\dagger\hat{\sigma}^j_- )$, where $\hat{\sigma}^j_-$ and $\hat{\sigma}^j_+$ are the spin lowering and raising operators. The ensemble coupling rate to the cavity is given by $\gens^2=\int p(g)g^2 dg$. The derivations that follow do not require any assumption on the shape of the distribution $p(g)$. We however assume that the spins are coupled to the electromagnetic field only through the cavity, an hypothesis that is very well verified for spin-cavity system at microwave frequencies. We consider the spin spectral distribution $n$ to be a Lorentzian centered at the mean spin Larmor frequency $\omega_s$, \ie $n(\Delta_j=\omega_j-\omega_s) = \frac{\Gamma}{2 \pi} \frac{1}{\frac{\Gamma^{2}}{4}+\Delta_j^{2}},$ and uncorrelated with the spin coupling distribution $p(g_j).$ All spins are assumed to be in their ground state at the beginning of the protocol.

We consider a simple echo-based protocol proposed in earlier works~\cite{Afzelius_2013} illustrated in \cref{fig:1}\textbf{(b)}.  During the first stage of the protocol, an incoming signal of envelope $\Ein(t)$, at mean Larmor frequency $\omega_s$, is absorbed by the spin ensemble. Two refocusing pulses applied at $t= \tau_{\text{a}}$ and $2\tau_{\text{a}}+\tau_{\text{e}}$ let us retrieve this absorbed quantum state at time $T_{\text{E}}=2\tau_{\text{a}}+2\tau_{\text{e}}$. The echo occurring between the two refocusing pulses, at time $2\tau_a$, is suppressed by detuning the cavity between the refocusing pulses $\tau_{\text{a}} < t < 2\tau_{\text{a}}+\tau_{\text{e}}$ \cite{ranjan2022}. During absorption and emission, the equations of motion (EOMs) of the system in the frame rotating at $\omega_s$, read~\cite{Afzelius_2013}
\begin{align}
    \frac{d}{d t} \varepsilon =&-\frac{\kappa_{0}+\kappa(t)+2 i \Delta_{\text{cs}}}{2} \varepsilon \nonumber \\
    &+i \sum_{j=1}^N g_j \sigma^{j}_-+\sqrt{\kappa(t)} \Ein(t), \label{eq:AbsE} \\
\frac{d}{d t} \sigma^{j}_- =&-i \Delta_j \sigma^{j}_-+ i g_j \varepsilon, \label{eq:AbsSigma}
\end{align}
where $\sigma_-^{j}(t)$ and $\varepsilon(t)$ are the expectation values of the operators $\hat\sigma_-^{j}$ and $\hat{\varepsilon}$, while $\Delta_{\text{cs}}=\omega_c-\omega_s$ is the detuning between the cavity and the spin central frequency. During emission, there is no input field, \ie the envelope $\Ein(t)=0$. In the above EOMs, we include intrinsic losses for the cavity field without considering any population or phase decays of individual spins, as their coherence time is assumed to be much longer than the duration of the absorption or emission steps. We also do not consider any spin-spin interaction. Finally, in \cref{eq:AbsSigma}, we have already implemented a mean-field approximation. Indeed, its last term should write as $ -i g_j \langle \hat{\sigma}_z^{j} \hat{\varepsilon} \rangle $.  It can be simplified through the Holstein-Primakoff approximation~\cite{Holstein-Primakoff,Afzelius_2013,julsgaard} to a factorized product $- i g_j\langle \hat{\sigma}_z^j \rangle \langle \hat{\varepsilon} \rangle$. Since we are considering the incoming field $\Ein$ only carries a few photons compared to the large number of spins ($N > 10^4$) in the ensemble, we expect $\langle \hat{\sigma}_z^j \rangle\sim-1$ during absorption. This mean population is also recovered after the two refocusing pulses, yielding the simplified form of \cref{eq:AbsSigma} we are using. 

\section{Absorption and emission}
\subsection{Quantum cascaded model}
We now cast the interaction of the cavity with the spin ensemble as an interaction with a bosonic mode $\Sigma$ coupled to a communication channel. To introduce this bosonic mode, we replace the term $\sum_{j=1}^N g_j \sigma^{j}_-$ in \cref{eq:AbsE} with its integrated version $\int_g \int_\Delta  g \sigma_-^{\Delta,g} p(g) n(\Delta)  dg d\Delta.$ 
By integrating the EOM of $\sigma_-^{j}$  (\cref{eq:AbsSigma}) during the absorption step, we can proceed to the elimination of $\sigma_-^{\Delta,g}$ in the above mentioned term, so that it becomes solely dependent on $\varepsilon_{\text{a}}$, the cavity mean field during the absorption step (see \cref{app:EOMs}, \cref{eq:spinsApp})
\begin{equation}
\begin{split}
    \sum_{j=1}^N g_j \sigma^{j,\text{a}}_- \rightarrow    i \gens^{2}   \int_{-\infty}^{t} d s\varepsilon_{\text{a}}(s)  \int d \Delta n(\Delta) e^{-i \Delta(t-s)}.
\end{split}
\end{equation}
By recognizing the Fourier transform of the Lorentzian function $n(\Delta)$, we can equalize this term to $i\gens \Sigma_{\text{a}}(t)$, where
$ \Sigma_{\text{a}}(t)= \gens \int_{-\infty}^{t} ds  e^{-\frac{\Gamma}{2}(t-s)} \varepsilon_{\text{a}}(s)$,
with $\Sigma_{\text{a}}(-\infty)=0.$ The variable $\Sigma_{\text{a}}$ can also be defined by an EOM:
\begin{equation}
\frac{d}{d t} \Sigma_{\text{a}} = -\frac{\Gamma}{2} \Sigma_{\text{a}} + \gens \varepsilon_{\text{a}}(t).
\label{eqn:defSigma1}
\end{equation}
We can now rewrite the intra-cavity EOM (\cref{eq:AbsE}) using $\Sigma_{\text{a}}$, and we obtain a new expression for the dynamics of the cavity
\begin{equation}
\begin{split}
\frac{d}{d t} \varepsilon_{\text{a}} =-\frac{\kappa_{0}+\kappa_{\text{a}}(t)+i2\Delta_{\text{cs}}}{2} \varepsilon_{\text{a}} &\\
- \gens & \Sigma_{\text{a}}+\sqrt{\kappa_{\text{a}}(t)} \Ein(t).
\end{split}
\label{eq:Abs2a}
\end{equation}
~\cref{eq:Abs2a,eqn:defSigma1} form a closed set of equations describing the full dynamics of the spin-cavity system during absorption $t<\tau_{\text{a}}$. We can also recognize that they correspond to the evolution of the mean-values of two bosonic modes, $\hat{\varepsilon}_{\text{a}}$ and $\hat{\Sigma}_{\text{a}}$, which interact with strength $\gens$. Moreover, we can identify the decay rates for each mode: respectively $\kappa_0$ and $\kappa_{\text{a}}$ for the cavity field $\varepsilon_{\text{a}}$, and $\Gamma$ for the mode $\Sigma_{\text{a}}$. This decay corresponds to the emission of a fictitious field $\Sigma_{\text{a,out}}=\sqrt{\Gamma}\Sigma_{\text{a}}$ into a newly defined spin channel, see~Fig.~\ref{fig:1}\textbf{(c)}. One can intuitively understand the nature of $\Sigma_{\text{a}}$ as the field stored in a collective “bright" mode of the spin ensemble, which is maximally coupled to the cavity. However, due to the ensemble broadening, the energy stored in this bright mode rapidly dissipates into “dark" modes, representing the uncoupled degrees of freedom of the spin ensemble. This process corresponds in our model to emission into the spin channel.

Refocusing techniques, acting as time reversal, permit to recover the initial magnetization created on the spin ensemble. The resulting filtered field will then drive the two bosonic modes system during emission, in a quantum cascade arrangement.
To highlight this quantum cascade, we introduce a new cavity variable $\varepsilon_{\text{e}}$ and new spin variables $\sigma_-^{j,\text{e}}$ to describe the system during the emission step ($t>2\tau_{\text{a}}+\tau_{\text{e}}$), obeying respectively the EOMs \cref{eq:AbsE} (without the driving term) and \cref{eq:AbsSigma}. 
Similarly to absorption, we can integrate the EOM of $\sigma_-^{j,\text{e}}$ to obtain an expression for the term $\sum g_j \sigma_-^{j,\text{e}} $ that solely depends on $\varepsilon_{\text{a}}$ and $\varepsilon_{\text{e}}$ (see \cref{app:EOMs}, \cref{eq:spinsEm}), so that the cavity field evolution is given by 
\begin{equation}
    \frac{d}{d t} \varepsilon_{\text{e}}  = -\frac{\kappa_{0}+\kappa_{\text{e}}(t)+i2\Delta_{\text{cs}}}{2} \varepsilon_{\text{e}}  - \gens \Sigma_{\text{e}} ,
    \label{eq:Em2}
\end{equation}
where $\kappa_{\text{e}}(t)$ is the cavity coupling rate during the emission part. Similarly to the absorption step, we can recognize that the cavity mode interacts at strength $g_{\mathrm{ens}}$ with the bosonic mode $\Sigma_{\text{e}}$. As detailed in \cref{app:EOMs}, we find that this bosonic mode is defined by the following EOM
\begin{equation}
    \frac{d}{d t} \Sigma_{\text{e}}=-\frac{\Gamma}{2}\Sigma_{\text{e}}+\gens \varepsilon_{\text{e}}(t)+\sqrt{\Gamma} \Sigma_{\text{e,in}},
    \label{eq:defSigma2}
\end{equation}
assuming $\Sigma_{\text{e}}(-\infty)=\varepsilon_{\text{e}}(-\infty)=0$. Here, $\Sigma_{\text{e,in}}$ corresponds to a drive field for this new spin bosonic mode. It corresponds to a filtered version of $\Sigma_{\text{a,out}}$ given by $\Sigma_{\text{e,in}}=\sqrt{\Gamma}\tilde{\Sigma}_{\text{a}}(T_{\text{E}}-t)$, where $\tilde{\Sigma}_{\text{a}}$ is defined by the following EOM (see~Appendix~\ref{app:EOMs})
\begin{equation}
    \frac{d}{d t} \tilde{\Sigma}_{\text{a}} =-\frac{\Gamma}{2}\tilde{\Sigma}_{\text{a}}+\gens \varepsilon_{\text{a}}(-t),
    \label{eqn:defSigma1tilde}
\end{equation}
with $\tilde{\Sigma}_{\text{a}}(-\infty)=0$. By exploiting the symmetry in ~\cref{eqn:defSigma1,eqn:defSigma1tilde}, we can express the filter function $H$ between the emitted spin field during absorption $\Sigma_{\text{a,out}}$, and the incoming spin drive field during emission $\Sigma_{\text{e,in}}$ as
\begin{equation}
    H[\delta]=\frac{\Sigma_{\text{e,in}}[\delta]}{\Sigma_{\text{a,out}}[\delta]}=\frac{\frac{\Gamma}{2}+i\delta}{\frac{\Gamma}{2}-i\delta}e^{-i \delta T_{\text{E}}}.
    \label{eq:filterFunc}
\end{equation}

\begin{figure}
    \centering
    \includegraphics[width=\linewidth]{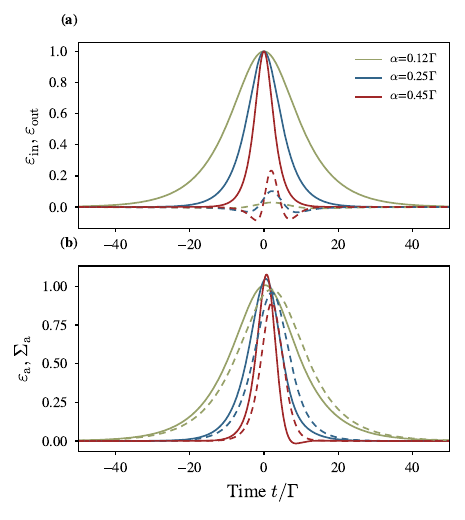}
    \caption{Response of the coupled spin-cavity system during the absorption step in case of a cosecant input pulse $\sqrt{a/2}/\cosh(\alpha t)$ of increasing bandwidth $\alpha$ derived through \cref{eqn:defSigma1,eq:Abs2a} with a uniform cavity coupling rate $\kappa_a$ chosen to match the cavity-spin system to 1 ($4\gens^2/(\Gamma\kappa_a) =1$). We take $\gens/\Gamma=0.5$ and $\kappa_0/\Gamma = 1/30$. \textbf{(a)} Input (solid) and reflected (dashes) cavity field. \textbf{(b)} Intracavity (solid) and spin (dashes) fields.}
    \label{fig:1bis}
\end{figure}

Using \cref{eq:Abs2a,eqn:defSigma1,eq:Em2,eq:defSigma2}, we identify a quantum cascaded system \cite{Carmichael_93,Gardiner93} represented in \cref{fig:1}\textbf{(c)}: the two interacting bosonic modes $\varepsilon_{\text{a}}$ and $\Sigma_{\text{a}}$, driven by a cavity drive $\Ein$ and emitting a spin field $\Sigma_{\text{a,out}}$, govern the dynamics of the two interacting bosonic modes $\varepsilon_{\text{e}}$ and $\Sigma_{\text{e}}$ through the spin drive $\Sigma_{\text{e,in}}$. In Fig.~\ref{fig:1bis}, we compute the response of the system to an incoming pulse of various bandwidths. We show that a slow pulse is perfectly absorbed when ensuring the cavity is \textit{matched} to the spin ensemble ($4 \gens^2/(\kappa_{\mathrm{a}}\Gamma)=1$, while for a faster pulse, the reflected field is non-zero, indicating an imperfect absorption.

\subsection{Efficiency definition}
We define the efficiency of the absorption step as the ratio between the energy transferred from the drive field $\Ein$ to the field stored inside the spin ensemble, and the energy of that drive field
\begin{equation}
    \eta_{\text{a}}=\frac{\int dt |\Sigma_{\text{a,out}}(t)|^2}{\int dt |\Ein(t)|^2}.
    \label{eq:eff_abs}
\end{equation}
Similarly, the efficiency of the emission step is given by the ratio of the energy transferred from the spin field, $\Sigma_{\text{e,in}}$, to the energy of the field emitted by the cavity $\varepsilon_{\text{e,out}}=\sqrt{\kappa_{\text{e}}(t)}\varepsilon_{\text{e}}$, namely
\begin{equation}
    \eta_{\text{e}}= \frac{\int dt \ |\varepsilon_{\text{e,out}}(t)|^2}{\int dt \ |\Sigma_{\text{e,in}}(t)|^2},\label{eq:eff_em}
\end{equation}
where the integrals are understood to run over $t \in (-\infty,\infty)$ (\cref{app:EOMs}). The overall efficiency of the protocol is given by $\eta=\eta_{\text{a}} \times \eta_{\text{H}} \times \eta_{\text{e}}$, where $\eta_{\text{H}}$ corresponds to the filter effect given by the refocusing pulses \cref{eq:filterFunc}, \ie $\eta_{\text{H}}= \frac{\int dt \ |\Sigma_{\text{e,in}}(t)|^2}{\int dt |\Sigma_{\text{a,out}}(t)|^2}$. Here, we consider perfect refocusing pulses, namely $\eta_{\text{H}}=1$.

Next, in \cref{subsec:kconst}, we derive the absorption and emission efficiencies $\eta_{\text{a}}$ and $\eta_{\text{e}}$ for a slow-pulse limit of the quantum cascaded system, as closed-form expressions in terms of the physical parameters $\gens$, $\kappa_0$ and $\Gamma$. We will then show that these efficiencies represent an upper bound when considering faster pulses in \cref{sec:Optimization}.

\section{Memory efficiency in the continuous-drive limit}
\label{subsec:kconst}

In this section, we consider the quantum cascaded model driven by a continuous-wave drive, with a constant coupling rate $\kappa$. We are interested in this steady-state behavior to derive the optimal value of $\kappa$, as a function of the physical parameters $\gens$, $\Gamma$ and $\kappa_0$, which maximizes the energy transmitted to the spins and then remitted by refocusing, or equivalently the efficiency $\eta$ defined above. 
The limit we derive here is a “slow-pulse” limit, which corresponds to pulses whose bandwidth is much narrower than the system bandwidth. While this limit might not be very useful in practice, as the pulse length should be much shorter than the coherence time of the quantum system generating it, it will provide a useful upper bound for $\eta$ in the case of realistic finite-length pulses, as treated in the next \cref{sec:Optimization}.

We define the following transmission coefficients associated with absorption and emission 
\begin{equation}
\begin{split}
    t_{\text{a}}=\Sigma_{\text{a,out}}/\Ein  , \ t_{\text{e}}=\varepsilon_{\text{e,out}}/\Sigma_{\text{e,in}},
    \label{eq:trCoeff}
\end{split}
\end{equation}
with $t_{\text{a}}$ being the transmission coefficient from the driveline to the spins, while $t_{\text{e}}$ is the transmission coefficient from the spins back to the driveline. Note that these transmission coefficients are related to the transmissivities of the spin-cavity channel $|t_{\text{a}}|^2,|t_{\text{e}}|^2$ ~\cite{transmissivity}. Using the EOMs for absorption and emission written in the frequency domain, these transmission coefficients are, in the Fourier domain, (see \cref{app:coeffs})
\begin{align}
\begin{split}
t_{\text{a}}[\delta] &= \frac{2\sqrt{\ks \kappa_{\text{a}}}}{\kappa_0+\kappa_{\text{a}}+\tilde{\kappa}_{\text{s}} + 2 i (\tilde{\delta}+\Delta_{\text{cs}})},\\
t_{\text{e}}[\delta] &=-\frac{2\sqrt{\ks \kappa_{\text{e}}}}{\kappa_0+\kappa_{\text{e}}+\tilde{\kappa}_{\text{s}} + 2 i (\tilde{\delta}+\Delta_{\text{cs}})},
\label{eq:tea}
\end{split}
\end{align}
with frequency dependence given by $\tilde{\kappa}_{\text{s}}/\ks = 1-\delta(\delta+\Delta_{\text{cs}})/\gens^2$ and $\tilde{\delta}/\delta = 1+ (\kappa_0+\kappa_{\text{a/e}})/\Gamma$. 
In absence of modulation, with constant coupling rate to the input line, the absorption and emission are described by a reciprocal linear system, so that the transmission coefficients $|t_{\text{a}}|$ and $|t_{\text{e}}|$ are symmetric under an interchange $\text{a} \leftrightarrow \text{e}$.

The efficiencies defined in \cref{eq:eff_abs,eq:eff_em} can then be rewritten as a function of the transmission coefficients, by making use of \cref{eq:trCoeff} and of Parseval's theorem to pass to the frequency domain
\begin{equation}
    \eta_{\text{a}}=\frac{\int d\delta |t_{\text{a}}[\delta] \Ein[\delta]|^2}{\int d\delta |\Ein[\delta]|^2}, \ \eta_{\text{e}}= \frac{\int d\delta |t_{\text{e}}[\delta] \Sigma_{\text{e,in}}[\delta]|^2}{\int d\delta |\Sigma_{\text{e,in}}[\delta]|^2}.
    \label{eq:effsFreq}
\end{equation}
From these expressions for the efficiencies, it is clear that an optimization of the memory protocol comes from a maximization of the transmissivities $|t_{\text{a}}|^2$, and $|t_{\text{e}}|^2$. To this end, we first maximize the transmissivities, as obtained from \cref{eq:tea}, at each frequency value $\delta$, with respect to $\Delta_{\text{cs}},$ finding the optimal $\Delta_{\text{cs}}^*[\delta]=\left(\frac{\ks/\Gamma}{1+4(\delta/\Gamma)^2}-1\right)\delta.$ Then, we evaluate the transmissivity at $\Delta_{\text{cs}}^*$, and maximize it with respect to $\kappa_{\text{a/e}}$, to obtain equal values for emission and absorption $\kappa_{\text{a/e}}^*[\delta]=\kappa_0+\frac{\ks}{1+4(\delta/\Gamma)^2},$ leading then to equal optimal transmissivities $|t_{\text{a/e}}[\delta]|^2=\frac{\ks}{\ks+\kappa_0+4\kappa_0(\delta/\Gamma)^2}$. Finally,  these are clearly maximized for $\delta^*=0$. 
We thus find that the sweet spot for the quantum memory protocol occurs at zero frequency, \ie for constant input $\Ein(t)=\Ein$, leading to zero detuning $\Delta_{\text{cs}}^*=0$, and
$\kappa_{\text{a/e}}^*=\kappa_0+\ks$, a coupling rate that maximizes both the transmission from the driveline to the spins and the transmission from the spins back to the driveline. 

\begin{figure}
    \centering
    \includegraphics[width=\linewidth]{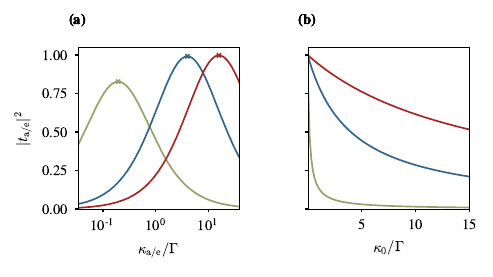}
    \caption{Transmissivity in the continuous-drive limit, when the cavity, the spins and the drive are all at resonance, as a function of cavity-driveline coupling (\textbf{(a)}), and of internal losses (\textbf{(b)}), for three different values of $\gens/\Gamma$ ($0.2$ green, $1$ blue, $2$ red curve). In \textbf{(a)} the internal losses decay rate is fixed to $\kappa_0/\Gamma=1/30$. The crosses correspond to the optimal coupling $\kappa_{\text{a/e}}^*=\kappa_0+\ks$, which indeed maximizes the transmissivity. In \textbf{(b)} the coupling is set to its optimal value $\kappa_{\text{a/e}}^*$, showing that the detrimental effect of intrinsic losses can be mitigated by increasing the coupling between the spins and the cavity.}
    \label{fig:2}
\end{figure}

The transmission coefficients at this sweet spot are then constant in time and read
\begin{equation}
    t_{\text{a}}=-t_{\text{e}}=\sqrt{\frac{\ks}{ \ks+\kappa_0}},
\end{equation}
while the protocol efficiency is
\begin{equation}
\eta=\eta_{\text{a}}^2=|t_{\text{a}}|^4=\left(\frac{\ks}{\kappa_0+\ks}\right)^2.
    \label{eq:effKappaconst}
\end{equation}
This expression highlights that the quality of the memory is given by the ratio $\ks/\kappa_0$, which describes how well the cavity is coupled to the spins compared to its coupling to the environment.  \cref{fig:2} shows the transmissivity at zero detuning as a function of the coupling rate $\kappa_{\text{a/e}}$ (panel \textbf{(a)}) and of internal losses $\kappa_0$ (panel \textbf{(b)}) for increasing values of $\gens/\Gamma$.
In panel \textbf{(a)} the crosses mark the sweet spot $\kappa_{\text{a/e}}^*=\kappa_0+\ks$ at $\kappa_0/\Gamma=1/30,$ while the transmissivity depicted in panel \textbf{(b)} is evaluated at the sweet spot $\kappa_{\text{a/e}}^*$.
Moreover, panel \textbf{(b)} shows that even in this case of optimal energy transfer, the cavity internal losses have a detrimental effect on the transmissivity, which can be overcome by increasing the coupling rate between the spins and the cavity, \ie $\ks.$ The gain in transmissivity for larger $\ks$ is also clear in the other panel.

Along with the transmitted energy comes the reflected energy, which is associated with the reflection coefficients $r_{\text{a}} = \varepsilon_{\text{a,out}}/\Ein$ back to the driveline and $r_{\text{e}}=\Sigma_{\text{e,out}}/\Sigma_{\text{e,in}}$ back to the spins. Their expressions are presented in \cref{app:coeffs} (see \cref{eq:appcoeffAbs} and \cref{eq:appcoeffEm}), and if we evaluate them at the transmission sweet spot ($\kappa_{\text{a/e}}^*$, $\delta^*$ and $\Delta_{\text{cs}}^*$) we find $r_{\text{a}}=0$, \ie the incoming drive is perfectly absorbed and $r_{\text{e}}=\frac{\kappa_0}{\kappa_0+\ks}$. This means that even in the case of optimal energy transfer, some energy remains within the spins during emission.

At the optimal operation point for constant coupling rate, the cooperativity of the system is $C^*=\frac{4 \gens^2}{\Gamma (\kappa_0+\kappa_{\text{a/e}}^*)}=\frac{\ks}{(2\kappa_0+\ks)}\leq1$, giving the efficiency of the protocol as
\begin{equation}
    \eta = \left(\frac{2C^*}{C^*+1}\right)^2.
\end{equation}
Unit efficiency is thus achieved under unit cooperativity (the so-called impedance matching condition~\cite{Afzelius_2013}), whenever the intrinsic losses of the cavity can be neglected.

\section{Optimal memory efficiency for temporal pulses}
\label{sec:Optimization}
We now wish to derive the temporal modulation of the coupling strengths during absorption, $\kappa_{\text{a}}(t)$, and emission, $\kappa_{\text{e}}(t)$, which optimize the storage efficiency of the memory $\eta=\eta_{\text{a}}\times \eta_{\text{e}}$ for an incoming pulse $\Ein(t)$. Based on the continuous-drive limit we derived above, we assume that the optimal spin-cavity detuning remains similar, and we consider $\Delta_{\text{cs}}=0$ in this section. Consequently, all the field variables are real.
Before describing the optimization problem, we perform a time rescaling $\tau=\Gamma t$: $\bar \delta=\delta/\Gamma$,  $\bar{\kappa}_{0}=\kappa_0/\Gamma$, 
$\bar{\kappa}_{\text{s}}=\ks/\Gamma$, 
$ \mathfrak{g}=\gens/\Gamma$, 
$\bar{\kappa}(\tau)=\kappa(\tau/\Gamma)/\Gamma$, 
$\bar{T}_{\text{E}}=\Gamma T_{\text{E}}$. The resulting equations are dimensionless and describe the memory protocol dynamics in units of the spin decay rate $\Gamma$, which sets a natural time scale for storage into the memory system. 
We denote the cavity and spin fields $\varepsilon_{\text{a/e}}(\tau/\Gamma)$ and $\Sigma_{\text{a/e}}(\tau/\Gamma)$, as well as the input drive fields for absorption $\Ein(\tau/\Gamma)$ and emission $\tilde\Sigma_a(\tau/\Gamma)$, all functions of the rescaled time $\tau$, by $\mathcal{E}_{\text{a/e}}(\tau)$, $\mathcal{S}_{\text{a/e}}(\tau)$, and $\sqrt{\Gamma}\mathcal{E}_{\text{in}}(\tau)$, $\mathcal{S}_{\text{in}}(\tau)$. With these
notations, the EOMs describing emission write
\begin{equation}
\begin{split}
\frac{d}{d \tau} \mathcal{E}_{\text{e}}  &= -\frac{\bar{\kappa}_{0}+\bar{\kappa}_{\text{e}}(\tau)}{2} \mathcal{E}_{\text{e}}  - \mathfrak{g} \mathcal{S}_{\text{e}},\\
\frac{d}{d \tau}  \mathcal{S}_{\text{e}} &=-\frac{1}{2}\mathcal{S}_{\text{e}}+ \bargens \mathcal{E}_{\text{e}}+ \mathcal{S}_{\text{in}}(\bar T_E-\tau),\\
\frac{d}{d \tau}  \mathcal{S}_{\text{in}} &= -\frac{1}{2}\mathcal{S}_{\text{in}}+\mathfrak{g}  \mathcal{E}_{\text{a}}(-\tau),
\end{split}
\label{eq:Emsren}
\end{equation}
while the EOMs for the absorption process are
\begin{equation}
\begin{split}
\frac{d}{d \tau} \mathcal{E}_{\text{a}} & =-\frac{\bar{\kappa}_{0}+\bar{\kappa}_{\text{a}}(\tau)}{2} \mathcal{E}_{\text{a}}- \mathfrak{g} \mathcal{S}_{\text{a}}+\sqrt{\bar{\kappa}_{\text{a}}(\tau)} \mathcal{E}_{\text{in}}, \\
\frac{d}{d \tau} \mathcal{S}_{\text{a}} & = -\frac{1}{2} \mathcal{S}_{\text{a}} + \mathfrak{g} \mathcal{E}_{\text{a}}.
\end{split}
\label{eq:Absren}
\end{equation}


\subsection{Optimizing absorption}
\label{subsec:optA}

Instead of fixing the shape of the incoming pulse $\Eein(\tau)$, for simplicity, we equivalently fix the shape of $ \mathcal{S}_{\text{a}}(\tau)$, such that the energy emitted into the spin ensemble equals 1. In this way, the shapes of $\mathcal{E}_{\text{a}}$ and $\Eein$ can be found by the absorption EOMs (\ref{eq:Absren}) and can be expressed in terms of $ \mathcal{S}_{\text{a}}(\tau)$ and its derivatives $\dot{\mathcal{S}}_{\text{a}}(\tau)$ and $\ddot{\mathcal{S}}_{\text{a}}(\tau)$, as well as $\bar{\kappa}_{\text{a}}(\tau)$. This last function is the only remaining degree of freedom that we would like to determine by solving an optimization problem which maximizes the absorption efficiency $\eta_{\text{a}}=\frac{\int d\tau | \mathcal{S}_{\text{a}}(\tau)|^2}{\int d\tau |\Eein(\tau)|^2}.$

Specifically, this optimization problem can be formulated as follows. For a given $\mathcal{S}_{\text{a}}$ such that $\int d\tau |\mathcal{S}_{\text{a}}(\tau)|^2=1$, solve
\begin{equation}
    \min_{\bar{\kappa}_{\text{a}}(\tau)\geq 0} \int d\tau |\Eein(\tau)|^2.
    \label{eq:absPr}
\end{equation}
Noting that by \cref{eq:Absren} $\mathcal{E}_{\text{in}}=A[\mathcal{S}_{\text{a}},\dot{\mathcal{S}}_{\text{a}},\ddot{\mathcal{S}}_{\text{a}}]/
\sqrt{\bar\kappa_{\text{a}}} +\sqrt{\bar\kappa_{\text{a}}} B[\mathcal{S}_{\text{a}},\dot{\mathcal{S}}_{\text{a}},\ddot{\mathcal{S}}_{\text{a}}]$ with

\begin{align}
\begin{split}
A[\mathcal{S}_{\text{a}},\dot{\mathcal{S}}_{\text{a}},\ddot{\mathcal{S}}_{\text{a}}]=& \frac{1}{\bargens}(\ddot{\mathcal{S}}_{\text{a}}+\frac{1+\bar\kappa_0}{2}\dot{{\mathcal{S}}_{\text{a}}}+\frac{4\bargens^2+\bar\kappa_0}{4}{\mathcal{S}}_{\text{a}})\\
B[\mathcal{S}_{\text{a}},\dot{\mathcal{S}}_{\text{a}},\ddot{\mathcal{S}}_{\text{a}}]=&\frac{1}{2\bargens}(\dot{{\mathcal{S}}_{\text{a}}}+{\mathcal{S}}_{\text{a}}/2),
\label{eq:ABdefinition}
\end{split}
\end{align}
we can re-express the optimization problem as
\begin{equation}
    \min_{\bar{\kappa}_{\text{a}}(\tau)\geq 0} \int d\tau \left(\frac{A(\tau)}{\sqrt{\kappa_{\text{a}}(\tau)}}+B(\tau)\sqrt{\kappa_{\text{a}}(\tau)}\right)^2,
    \label{eq:absPr2}
\end{equation}
whose optimal solution is clearly given by
\begin{equation}\label{eq:kappa1star}
\bar \kappa_{\text{a}}^*(t)=\left|\frac{A(t)}{B(t)}\right|.
\end{equation}

Now, given the optimal solution $\bar{\kappa}_{\text{a}}^*$, depending on the choice of $\mathcal{S}_{\text{a}}$, we aim to derive an upper bound on the absorption efficiency, that is, a lower bound for the functional in~\eqref{eq:absPr}. Using the functions $A(t)$ and $B(t)$ decomposing $\bar{\kappa}_{\text{a}}^*$, we can express the optimal value of the functional as
\begin{equation}
    \min \int |\Eein(\tau)|^2 d\tau = \int \left| \frac{B}{A}\right| \left(A+ B \left| \frac{A}{B}\right|\right)^2 d\tau.
    \label{eq:Emin}
\end{equation}
The integrand $\left| \frac{B}{A}\right| \left(A+ B \left| \frac{A}{B}\right|\right)^2$ can be rewritten as $2|AB|+2 AB$ which is lower bounded by $4AB$. Using the definitions of $A$ and $B$ in \cref{eq:ABdefinition}, we can show that (see \cref{app:inOutEnergies})
\begin{equation}
\begin{split}
       \min \int |\Eein(\tau)|^2 d\tau \geq & \\ \int d\tau & \left[4\frac{\bar{\kappa}_0}{\bar{\kappa}_{\text{s}}} \dot{\mathcal{S}}_{\text{a}}^2 +\left(1 +\frac{\bar{\kappa}_0}{\bar{\kappa}_{\text{s}}}\right) \mathcal{S}_{\text{a}}^2\right]. 
\end{split}
    \label{eq:disEin}
\end{equation}
Using Parseval's theorem, this inequality is equivalent to
\begin{equation}
\begin{split}
  \min \int |\Eein(\tau)|^2 d\tau \geq &\\
  \frac{1}{2\pi}& \int \frac{\bar{\kappa}_{\text{s}}+\bar\kappa_0(1+4\bar{\delta}^2)}{\bar{\kappa}_{\text{s}}}  | \mathcal{S}_{\text{a}}[\bar{\delta}]|^2 d\bar{\delta},  
\end{split}
\label{eq:inEnergy}  
\end{equation}
which is lower bounded by $\frac{\ks+\kappa_0}{2\pi\ks}\int d\bar{\delta} | \mathcal{S}_{\text{a}}[\bar{\delta}]|^2$. We recall that we have assumed $\frac{1}{2\pi} \int d\bar{\delta} | \mathcal{S}_{\text{a}}[\bar{\delta}]|^2=1,$ which implies that the absorption efficiency $\eta_\text{a}$ is directly upper-bounded by
\begin{equation}
    \eta_{\text{a}} \leq \frac{\ks}{\kappa_0+\ks}.
    \label{eq:absLB}
\end{equation}
As seen before, this upper bound corresponds to the maximal absorption efficiency \cref{eq:effKappaconst} found in the case of constant $\kappa_{\text{a}}$ for an adiabatic incoming pulse.

\subsection{Optimizing emission}
Now that we have found the profile of $\bar{\kappa}_{\text{a}}$ that realizes the maximal transfer of energy from the driveline to the spins, we aim to find the profile of $\bar{\kappa}_{\text{e}}$ that permits the maximal transfer of energy from the spins back to the driveline, namely we want to maximize $\eta_{\text{e}}=\frac{\int d\tau \bar{\kappa}_{\text{e}}(\tau) |\mathcal{E}_{\text{e}}(\tau)|^2}{\int d\tau  |\mathcal{S}_{\text{in}}(\tau)|^2}.$

To this end, we first express the denominator $\int d\tau  |\mathcal{S}_{\text{in}}(\tau)|^2$ using our knowledge of the absorption step. Through \cref{eq:filterFunc}, we can show that $\mathcal{S}_{\text{in}}(\bar{\delta}) = H[\Gamma\bar{\delta}] \mathcal{S}_{\text{a}}(\bar{\delta})$. Given that we have assumed perfect refocusing pulses, \ie $|H|=1$, we have $\int d\tau  |\mathcal{S}_{\text{in}}(\tau)|^2 =\int d\tau  |\mathcal{S}_{\text{a}}(\tau)|^2 =1$. Similarly to the absorption problem, maximizing the emission efficiency thus reduces to maximizing $E_{\text{out}}=\int d\tau \bar{\kappa}_{\text{e}}(\tau) \mathcal{E}_{\text{e}}(\tau)^2$.

Rewriting the second  \cref{eq:Emsren} as $\bargens\mathcal{E}_{\text{e}}=\dot{\mathcal{S}}_{\text{e}}+\frac{1}{2}\mathcal{S}_{\text{e}} - \mathcal{S}_{\text{in}},$ and
manipulating \cref{eq:Emsren} as shown in \cref{app:inOutEnergies}, it is possible to eliminate the dependence of $E_{\text{out}}$ on $\bar{\kappa}_{\text{e}},$ so that $E_{\text{out}}$ is simply expressed as a function of $\mathcal{S}_{\text{e}}$, its derivatives, and $\mathcal{S}_{\text{in}}$
\begin{equation}
    \begin{split}
        E_{\text{out}} = \int d\tau [\mathcal{S}_{\text{in}}^2 - (\mathcal{S}_{\text{e}}-\mathcal{S}_{\text{in}})^2 
& \\ -  \frac{\bar\kappa_0}{\bar{\kappa}_{\text{s}}} (\mathcal{S}_{\text{e}}+2\dot{\mathcal{S}_{\text{e}}}  & -2\mathcal{S}_{\text{in}})^2].
    \end{split}
    \label{eq:output_en}
\end{equation}
Furthermore, note that, following~\cref{eq:Emsren}, we have
\begin{align}
\bar{\kappa}_{\text{e}}(\tau)=-\bar{\kappa}_0-2 \frac{\ddot{\mathcal{S}}_{\text{e}}+1/2\dot{\mathcal{S}}_{\text{e}}-\dot{\mathcal{S}}_{\text{in}}+\tilde g^2 \mathcal{S}_{\text{e}}}{\dot{\mathcal{S}}_{\text{e}}+1/2 \mathcal{S}_{\text{e}}-\mathcal{S}_{\text{in}}}.
\end{align}
Therefore, the optimization problem can be written as a maximization of the output energy~\eqref{eq:output_en} as a function of $\mathcal{S}_{\text{e}}$, under the constraint
\begin{equation}\label{eq:em_const}
\bar{\kappa}_0+2 \frac{\ddot{\mathcal{S}}_{\text{e}}+1/2\dot{\mathcal{S}}_{\text{e}}-\dot{\mathcal{S}}_{\text{in}}+\tilde g^2 \mathcal{S}_{\text{e}}}{\dot{\mathcal{S}}_{\text{e}}+1/2 \mathcal{S}_{\text{e}}-\mathcal{S}_{\text{in}}}\leq 0.
\end{equation}
This constraint can be written in the following form
\begin{multline}\label{eq:constraint2}
\mathcal{S}_{\text{e}}\in \{\mathcal{S}_{\text{e}}~|h_1(\mathcal{S}_{\text{e}})\leq c_1(\tau),h_2(\mathcal{S}_{\text{e}})\geq c_2(\tau)\}\cup \\ \{ \mathcal{S}_{\text{e}}~|h_1(\mathcal{S}_{\text{e}})\geq c_1(\tau),h_2(\mathcal{S}_{\text{e}})\leq c_2(\tau)\}
\end{multline}
with
\begin{align*}
h_1(\mathcal{S}_{\text{e}})&=4\ddot{\mathcal{S}_{\text{e}}}+2(1+\bar\kappa_0)\dot{\mathcal{S}_{\text{e}}}+(\bar{\kappa}_{\text{s}}+\bar\kappa_0)\mathcal{S}_{\text{e}},\\
h_2(\mathcal{S}_{\text{e}})&=2\dot{\mathcal{S}_{\text{e}}}+\mathcal{S}_\text{e},\\
c_1(\tau)&=4\dot{\mathcal{S}_{\text{in}}}(\tau)+2\bar\kappa_0\mathcal{S}_{\text{in}}(\tau),\\
c_2(\tau)&=2\mathcal{S}_{\text{in}}(\tau).
\end{align*}
Therefore, the constraint can be seen as the union of two convex sets. It is possible to solve each of these convex optimization problems under constraints using an appropriate iterative algorithm, such as the Uzawa algorithm for finding the saddle point of the associated Lagrangian, and then compare the maximal values in each set to find the global maximum under constraints.

Here, for simplicity, we have decided to relax the constraint, find the unconstrained optimal output energy, and finally check if this optimal solution actually satisfies the constraint.
To this end, using Parseval's theorem, we re-express  the output energy in \cref{eq:output_en} as
\begin{equation}
    \begin{split}
          E_{\text{out}} = \frac{1}{2 \pi}\int d\bar{\delta} \Big[&\left|\mathcal{S}_{\text{in}}[\bar{\delta}]\right|^2 - \left|\mathcal{S}_{\text{e}}[\bar{\delta}]-\mathcal{S}_{\text{in}}[\bar{\delta}]\right|^2  
          \\
          -\frac{\bar\kappa_0}{\bar{\kappa}_{\text{s}}}  & \left|(1+2i\bar{\delta})\mathcal{S}_{\text{e}}[\bar{\delta}]-2\mathcal{S}_{\text{in}}[\bar{\delta}]\right|^2\Big] . 
    \end{split}
    \label{eq:outEnergy} 
\end{equation}
The corresponding optimal spin field (without constraint) is given by
\begin{equation}
    \mathcal{S}_{\text{e}}^*[\bar\delta]= \frac{2(1-2i\bar\delta)\bar\kappa_0+\bar{\kappa}_{\text{s}}}{(1+4\bar\delta^2)\bar\kappa_0+\bar{\kappa}_{\text{s}}}\mathcal{S}_{\text{in}}[\bar\delta].
    \label{eq:optSigma2}
\end{equation}
Using the second equation in (\ref{eq:Emsren}), we can also deduce the optimal cavity field  
\begin{equation}
\mathcal{E}_{\text{e}}^*=\sqrt{\bar{\kappa}_{\text{s}}}\frac{2i\bar\delta-1}{\bar\kappa_0(1+4\bar\delta^2)+\bar{\kappa}_{\text{s}}}\mathcal{S}_{\text{in}}[\bar\delta],
    \label{eq:optE2}
\end{equation}
while using the first \cref{eq:Emsren} we can determine the corresponding optimal cavity bandwidth modulation
\begin{equation}
        \bar{\kappa}_{\text{e}}^*(\tau) =-\bar{\kappa}_0-2 \frac{\mathcal{F}^{-1}[i\bar{\delta}\mathcal{E}_{\text{e}}^*+\bargens \mathcal{S}_{\text{e}}^*](\tau)}{\mathcal{F}^{-1}[\mathcal{E}_{\text{e}}^*](\tau)},
        \label{eq:kappa2star}
\end{equation}
where $\mathcal{F}^{-1}$ stands for the  inverse Fourier transform. 

In the absence of internal loss, $\kappa_0=0$, we find $\mathcal{S}^*_{\text{e}}=\mathcal{S}_{\text{in}}$, $\mathcal{E}_{\text{e}}^*=-\mathcal{E}_{\text{a}}(\tau-\bar T_E)$, corresponding to an emission efficiency of 1. However, we have to check if the corresponding cavity bandwidth modulation satisfies the positivity constraint, \ie
that $\bar\kappa_e^*$ given as follows is positive for all times (for $\kappa_0=0$)
\begin{align}
    \bar{\kappa}_{\text{e}}^*(\tau)&= \frac{4 \ddot{\mathcal{S}}_{\text{in}}-2 \dot{\mathcal{S}}_{\text{in}}+\bar{\kappa}_{\text{s}} \mathcal{S}_{\text{in}}}{\mathcal{S}_{\text{in}}-2\dot{\mathcal{S}}_{\text{in}}}.   
\end{align}

At this point, we again consider the case of non-vanishing internal loss $\bar\kappa_0>0$, and we derive an upper bound for the emission efficiency $\eta_{\text{e}}$. 
We plug in the general expression of $\mathcal{S}_{\text{e}}^*$ derived in \cref{eq:optSigma2} in the output energy (\ref{eq:outEnergy}), which yields
\begin{equation}
   E_{\text{out}}= \frac{1}{2\pi} \int d\bar{\delta} |\mathcal{S}_{\text{in}}|^2  \frac{\bar{\kappa}_{\text{s}}}{(1+4\bar{\delta}^2)\bar\kappa_0+\bar{\kappa}_{\text{s}}}.
   \label{eq:UpperBound0}
\end{equation}
This provides an upper bound for the emission efficiency as no positivity constraint on $\bar\kappa_e$ is imposed. This bound is actually saturated when \cref{eq:kappa2star} remains positive for every time $\tau$. 

Now, let us go even further and note that from the above equation we obtain
\begin{equation}
   E_{\text{out}}\leq\frac{\bar{\kappa}_{\text{s}}}{\bar{\kappa}_{\text{s}}+\bar\kappa_0} =\frac{\ks}{\ks+\kappa_0}
   \label{eq:UpperBound}
\end{equation}
with the inequality a consequence of $\frac{1}{2\pi} \int d\bar{\delta} |\mathcal{S}_{\text{in}}|^2 =1$.
We thus have  $\eta_{\text{e}} \leq \frac{\ks}{\kappa_0+\ks}$, which together with the previously derived bound (\ref{eq:absLB}) gives
\begin{equation}
    \eta = \eta_{\text{a}} \times \eta_{\text{e}} \leq  \left(\frac{\ks}{\kappa_0+\ks}\right)^2.
    \label{eq:ubound}
\end{equation}
As for absorption, this upper bound on absorption and re-emission efficiency is reached in the limit of adiabatic input drives and with a constant modulation $\kappa_\text{a/e}$ (see \cref{eq:effKappaconst}). This limit is entirely governed by the amount of cavity intrinsic losses compared to the spin-induced losses. In the following section, we perform numerical simulations to evaluate the efficiency cost in using pulses whose bandwidth lies beyond this adiabatic limit.

\section{Numerical simulations}
From \cref{sec:Optimization}, we can derive not only upper bounds on the efficiency but also determine the optimal modulation of the cavity coupling rates $\bar{\kappa}_{\text{a}}^*$ and $\bar{\kappa}_{\text{e}}^*$ (see \cref{eq:kappa1star,eq:kappa2star}), as well as the corresponding input drive $\Ein(t)$, given a fixed shape for the occupancy of the spin field mode $\Sigma_{\text{a}}$. By considering a particular wavepacket shape $u(\alpha \Gamma t)$, we can then derive the expected efficiency as a function of the wavepacket bandwidth $\alpha \Gamma$. Although this analysis does not allow us to identify an optimal pulse shape for a given bandwidth, it lets us quantify the reduction in efficiency in operating the quantum memory for this specific pulse shape given this bandwidth. 
From this study, we wish to extract the minimal wavepacket duration $\tau_{\mathrm{in}}^*$ beyond which a reduction in efficiency occurs, and compare it with two key metrics for a quantum memory. First, comparing  $\tau_{\mathrm{in}}^*$ to the memory storage time (given by the ensemble coherence time $T_2$) provides an estimate of the capacity of the memory, \ie the number of temporal modes it can potentially store. Second, in the context of a modular architecture, this wavepacket will realistically be emitted by a qubit. For a quantum memory to be pertinent in such an architecture, $\tau_{\mathrm{in}}^*$ must be short compared to the qubit lifetime to not limit the itinerant transfer between the qubit and the memory.
\subsection{Physical parameters}

We wish to realize this temporal analysis in the particular context of quantum memories operating at microwave frequencies that may eventually be coupled to superconducting quantum bits. The most promising spin candidates for these memories are, for example, donors in silicon~\cite{Wolfowicz2013} or rare-earth ions~\cite{tittel2025}. These spin systems are attractive candidates since they possess clock transition “sweetspots” where the decoherence created by spin-spin interactions is canceled~\cite{fraval2004}. For example, at these particular operating points, bismuth donors in silicon achieve $T_2^{\ce{Bi}}\approx \SI{1}{\second}$~\cite{Wolfowicz2013},
ytterbium ions in yttrium orthosilicate yield $T_2^{\ce{171Yb}:\ce{Y2SiO5}} = \SI{4}{\milli\second}$~\cite{alexander2022} and the same ions in scheelite $T_2^{\ce{171Yb}:\ce{CaWO4}} = \SI{0.15}{\second}$~\cite{tiranov2025}. Very strikingly, the last system exhibits this coherence time in the absence of a biasing magnetic field. These spin coherence times compare favorably to present-day record coherence time for superconducting qubits ($\approx\SI{1}{\milli\second}$), and even more favorably for qubits inserted in large-scale superconducting quantum processors or for qubits in quantum state transfer experiments where the achievable coherence time is rather about $\sim\SI{0.02}{\milli\second}$~\cite{Kurpiers2018,niu2023}.

Using an ensemble of these spins for implementing a microwave quantum memory will rely on patterning a superconducting microwave resonator implementing the cavity on top of the crystal containing the spins~\cite{RanjanPRL2020,O'SullivanPRX2022} or placing this crystal on top of another substrate on which the resonator is patterned~\cite{grezes2015}. In the former case, the strain induced by the superconducting film on the crystal can broaden the spin inhomogeneous linewidth (which may already be sizable) to $\Gamma/(2\pi)\sim\SI{10}{\mega\hertz}$, whereas in the latter geometry one may hope to be closer to its natural values, which can be as low as $\SI{5}{\kilo\hertz}$ for $\ce{^171Yb}:\ce{CaWO4}$ for instance~\cite{tiranov2025}. The particular choice of geometry for the superconducting resonator, combined with the particularities of the spin systems and the concentration of spins, sets the ensemble coupling constant~\cite{Grezes2014}. For example, donors in silicon samples cannot be too heavily doped, so will reach a more limited coupling constant $\gens/(2\pi)\sim\SI{100}{\kilo\hertz}$~\cite{RanjanPRL2020,osullivan2020} than for example some rare-earth ions where an ensemble coupling strength in the range of $\SI{4}{\mega\hertz}$ have been observed~\cite{staudt2012}. Overall, depending on the particular set of experimental conditions, the ratio $\gens/\Gamma$ can take values between $\num{0.02}$ and $200$. However, let us note that the larger ensemble coupling constants are typically reached using high-spin concentrations, which limit the coherence time achievable at the clock-transition through second-order spin-spin interactions~\cite{Wolfowicz2013}, so that a more reasonable upper bound for $\gens/\Gamma$ would be $\sim10$. These parameters correspond to spin-induced losses on the cavity of the order of $\ks/(2\pi) = \SI{1}{\kilo\hertz}$ to $\SI{10}{\mega\hertz}$. 

Let us now focus on the achievable cavity decay rates. For intrinsic losses, when the resonator is patterned on high-grade microwave substrates such as sapphire or silicon, decay rates as low as $\kappa_0/(2\pi)=\SI{10}{\kilo\hertz}$ may be achieved~\cite{kjaergaard2020a}. When using a spin-doped crystal made of other materials, more realistic values $\kappa_0/(2\pi)$ up to $\sim\SI{300}{\kilo\hertz}$ should be considered. The ratio $\kappa_0/\Gamma$ can thus take values from $\num{1e-4}$ up to 30. The derivation of \cref{subsec:kconst} makes clear that the coupling rate of the cavity to the input line should at minima match $\kappa_0 + \ks$. We assume in the following that this is the case and that no limitation will come from limited coupling to the measurement line. Experimentally, this would imply modulating $\kappa$ from less than $\SI{10}{\kilo\hertz}$ to more than $\SI{10}{\mega\hertz}$. This has already been achieved for superconducting resonators using Josephson junctions~\cite{kerckhoff2013a,grebel2024}, and can also be done on a smaller range with kinetic inductance when the resonator should be operated in a magnetic field~\cite{bockstiegel2016,wen2025}. Using these two techniques, modulating the resonator linewidth can be done with high precision for a broad range of desired shapes, with a response time of less than $\SI{10}{\nano \second}$.

\subsection{Wavepacket: hyperbolic secant case}
\label{subsec:hsPulse}

 \begin{figure*}
    \centering
    \includegraphics[width=\linewidth]{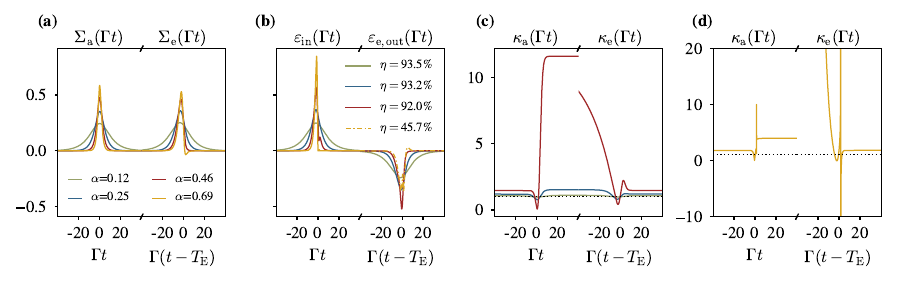}
    \caption{Numerical simulations results for a hyperbolic secant pulse, with system parameters $\gens/\Gamma=1/2$ and $\kappa_0/\Gamma=1/30,$ corresponding to critical speeds $\alpha_{\text{abs}}^*=0.474$ and $\alpha_{\text{em}}^*=0.5$.
    We compare a slow-adiabatic case ($\alpha=0.12$), with faster pulses just below ($\alpha=0.46$) and above $\alpha=0.69$ these thresholds. \textbf{(a)} Field stored in the spins during absorption and emission (dimensionless). \textbf{(b)} Incoming and retrieved field (in units of $\sqrt{\Gamma}$); evaluated using $\kappa_{\text{e}}(t)^*$ when defined, and $\kappa_{\text{a}}(T_E-t)$ when not (dashed-dot yellow curve).
    \textbf{(c)}-\textbf{(d)} Optimal modulation for cavity-driveline coupling during absorption (see \cref{eq:kappa1star}) and emission (see \cref{eq:kappa2star}) in units of $\Gamma$. As $\alpha$ is increased, getting away from the “slow-pulse” limit (black-dotted line), efficiency decreases (printed in \textbf{(b)}). At high speed ($\alpha>1/2$),  
    the denominator of \cref{eq:kappa1star} vanishes near $t=0$, creating a singularity in $\kappa_{\text{a}}(t)$.
    A similar divergence occurs for $\kappa_{\text{e}}(t)$ near $T_{\text{E}}$, accompanied by an unphysical modulation (crossing of the zero).
    }
    \label{fig:shapesComp}
\end{figure*}

To dictate the choice of the wavepacket $\Sigma_{\text{a}}(t)$ that we will consider in the rest of the study, we remark that the optimal shape for $\kappa_{\text{a}}$ is given by a ratio of the derivatives of $\Sigma_{\text{a}}(t)$, and by a ratio of Fourier transforms for $\kappa_{\text{e}}$. Using a wavepacket that is not at least twice differentiable, with a well-behaved Fourier transform, is thus not advised. In practice, we consider an hyperbolic secant pulse $\Sigma_{\text{a}}(t)=\sqrt{\alpha/2}/\cosh(\alpha \Gamma t)$ normalized to unity. In \cref{app:numerics}, we also perform the study for a Lorentzian-shaped wavepacket, which gives a higher requirement for the maximum coupling rate $\kappa_{\text{max}}$, and which performs worse in terms of efficiency compared to the hyperbolic secant pulse. The bandwidth of the input pulse is $\alpha \Gamma$, meaning that the spins are capable of absorbing all the frequency components of the input field within the ensemble linewidth $\Gamma$ if $\alpha \leq 1$.

To perform numerical simulations, we use this wavepacket to compute $\bar{\kappa}_{\text{a}}^*$ using \cref{eq:kappa1star}.  We then compute the input field $\mathcal{E}_{\text{in}}$ and the cavity field $\mathcal{E}_{\text{a}}$ through the EOMs (\ref{eq:Absren}).

For the emission process, we compute the dynamics in the frequency domain using \cref{eq:optSigma2,eq:optE2}, and then apply the inverse Fourier transform to obtain the time-domain evolution. The associated modulation of the coupling through \cref{eq:kappa2star} leads to maximal emission efficiency, but does not take into account the positivity constraint on $\bar{\kappa}_{\text{e}}^*(\tau)$.
When this constraint is violated, as occurs for pulse speeds $\alpha \geq \alpha_{\mathrm{em}}^*$, the output field
$\mathcal{E}_{\text{e,out}}=\sqrt{\bar{\kappa}_{\text{e}}}\mathcal{E}_{\text{e}}$ becomes ill-defined.
To address this issue in the fast-pulse regime ($\alpha \geq \alpha_{\mathrm{em}}^*$), where \cref{eq:kappa2star} yields negative values for the modulated coupling, we instead adopt a suboptimal but physical modulation: we mirror the absorption profile, setting $\kappa_{\text{e}}(t)=\kappa_{\text{a}}(T_{\text{E}}-t)$, which aligns with the time-reversed interpretation of emission in the quantum cascade formalism.
We then compute the cavity field $\mathcal{E}_{\text{e}}$, the spin field $\mathcal{S}_{\text{e}}$ through the EOMs (\ref{eq:Emsren}). The feedback term $\mathcal{S}_{\text{in}}$ is evaluated in the frequency domain as $\mathcal{S}_{\text{in}}(\bar{\delta}) = H[\Gamma\bar\delta] \mathcal{S}_{\text{a}}(\bar{\delta})$, then Fourier transformed back to the time domain. 
Finally, we compare the associated emission efficiency with the upper bound given by \cref{eq:UpperBound0}. This upper bound is saturated for pulse speeds
$\alpha < \alpha_{\text{em}}^*$.

Numerical results for the cavity and spin field, as well as the modulated coupling rates, are presented in \cref{fig:shapesComp} for four values of the bandwidth $\alpha=0.12, 0.25, 0.46$ and $0.69$. For this set of parameters, we find $\alpha_{\text{em}}^*=0.5$. In the small bandwidth limit ($\alpha=0.12$, green), the coupling rate approaches its “slow-pulse” optimal value $\kappa_0+\ks$ (black dotted line), and the numerically-determined efficiency ($\eta=93.5\%$) comes close to the adiabatic upper bound ($\eta=93.7\%$). Even at intermediate bandwidth ($\alpha=0.25$), there is little loss in efficiency ($\eta=93.2\%$). The input and outgoing cavity fields also closely resemble hyperbolic secant pulses. As we consider faster pulses, the optimal coupling modulations for absorption and emission lose in smoothness, and their minimum gets closer to 0. Similar to emission, we define the bandwidth $\alpha_{\text{abs}}^*$ at which a cancellation point appears in the optimal absorption coupling $\kappa_\text{a}^*(t)$. We observe that just below this limit ($\alpha=0.46$), the efficiency is still quite high ($\eta=92.0\%$). When switching to higher speed ($\alpha=0.69$, beyond even the definition region of $\bar{\kappa}_{\text{e}}^*$), the absorption efficiency is largely impacted ($\eta_{\text{abs}}=74.9\%$), evidencing a trade-off between the pulse bandwidth and its storage efficiency even though the optimal modulation function is still defined. We also see that the input pulse is considerably deformed and is rather akin to a simple exponential wavepacket. For emission, we observe that taking the suboptimal choice $\kappa_{\text{e}}(t)=\kappa_{\text{a}}(T_{\text{E}}-t)$ also leads to a rather limited efficiency ($\eta_{\text{em}}=61.0\%$), leading to a total efficiency of $\eta=45.7\%$. 


\subsection{Efficiency as a function of pulse bandwidth and intrinsic losses}

\begin{figure}[t!]
    \includegraphics[width=\linewidth]{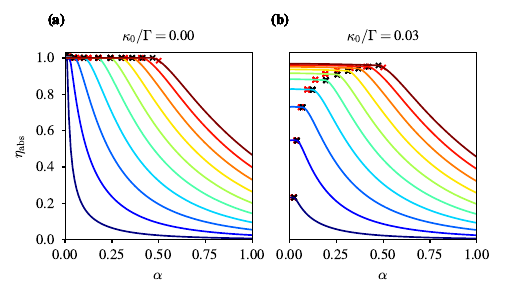}
    \includegraphics[width=0.55\linewidth]{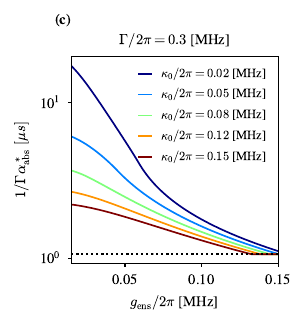}
    \centering
    \caption{Absorption efficiency analysis as a function of the pulse speed $\alpha$, increasing the ratio $\gens/\Gamma$ from $0.05$ (blue) to $0.5$ (dark red), for zero cavity internal losses \textbf{(a)} and when $\kappa_0/\Gamma=1/30$ \textbf{(b)}. When the input pulse is slow enough ($\alpha < \alpha^*_{\mathrm{abs}}$), the absorption efficiency is given by \cref{eq:etaAbsSlow}. Above this critical value $\alpha^*_{\mathrm{abs}}$ (black crosses), which increases with $\gens/\Gamma$, the efficiency drops. Its approximation by \cref{eq:alpha*Approx} is marked with red crosses. \textbf{(c)} Minimal pulse duration as a function of the spin-cavity coupling strength $\gens$, for different values of the cavity internal losses $\kappa_0/2 \pi$ (printed in the legend, in [MHz]). An increment of $\gens$ allows for a faster input pulse, up to a point set by the spins' bandwidth $\Gamma.$ A similar effect is observed when increasing $\kappa_0,$ as you can see from the relative decrease between the different curves.  We must recall that an increase of $\kappa_0$ corresponds as well to a decrease in efficiency, meaning that you are allowed to go faster but with an efficiency cost.
    }
    \label{fig:eta_vs_alpha}
\end{figure}

Restricting our analysis to hyperbolic secant pulses, we first study the efficiency of the absorption process as a function of the wavepacket bandwidth $\alpha\Gamma$ without (\cref{fig:eta_vs_alpha}\textbf{(a)}) and with intrinsic losses (\cref{fig:eta_vs_alpha}\textbf{(b)}).
In both cases, we observe that the efficiency lies close to the slow-pulse limit (\cref{eq:effKappaconst}) up to some critical bandwidth value after which the efficiency drops considerably.
We observe that the intrinsic losses, which irremediably lower the slow-pulse efficiency, seem to have a marginal effect on this bandwidth limitation. 
As we observed in \cref{fig:shapesComp}, this drop in efficiency ties to the optimal modulation shape derived in \cref{eq:kappa1star}, which expresses $\kappa_{\text{a}}^*(t)$ as the absolute value of the ratio of $A$ over $B$, where $A$ and $B$ are linear combinations of derivatives of the considered wavepacket. In our optimization problem, we only consider $A\neq0$ and $B\neq0$, which are non-zero at all times. At low speed, we find that $A/B$ is always positive, so that the absolute value does not play any role, and $A$ and $B$ are far from canceling at any point in time. 
In this case, \cref{eq:Emin} has a closed-form solution, which gives us an analytical expression of the absorption efficiency
\begin{equation}
    \eta_{\text{abs}}=\frac{\ks}{\ks+\kappa_0\left( 1+4\alpha^2/3\right)}
    \label{eq:etaAbsSlow}
\end{equation}
Noting that 
\begin{equation}
\begin{split}
        \kappa_{\mathrm{a}}^*(\tau)=\left|\kappa_0 +\Gamma  +\frac{\Gamma(1+4\alpha^2(2\operatorname{sech}(\alpha\tau)^2-1))-{\kappa}_s}{2\alpha \tanh(\alpha \tau)-1}\right|,
\end{split}
\end{equation}
The maximum limit for speed comes from having the denominator well-defined and non-zero at all times, which corresponds here to $\alpha<1/2$. Even below this limit, the nominator may present a cancellation point that would activate the absolute value. We find that after this critical speed $\alpha^*_{\mathrm{abs}}$ (represented by black crosses in \cref{fig:eta_vs_alpha}\textbf{(b)}), the optimal efficiency drops significantly due to this lack of a smooth control function, and \cref{eq:etaAbsSlow} gives just an upper bound for $\eta_{\text{abs}}$. 
For hyperbolic secant pulses, we can derive that the cancellation point appears for 
\begin{equation}
\alpha^*_{\mathrm{abs}} \approx \half \min\left[\frac{\ks + \kappa_0}{\Gamma+\kappa_0},1\right],
\label{eq:alpha*Approx}
\end{equation}
marked with red crosses in \cref{fig:eta_vs_alpha}\textbf{(b)}, which stand in the vicinity of the numerically evaluated $\alpha^*_{\mathrm{abs}}$ (black crosses).

In the absence of intrinsic losses on the cavity, this result implies that when $\ks>\Gamma$, the optimal bandwidth to operate at full efficiency is at best $\Gamma/2$. Otherwise said, the memory protocol is only efficient for pulses whose bandwidth is below $\Gamma/2$, or equivalently, whose temporal extent is longer than $2/\Gamma$. For smaller couplings, i.e. $\ks<\Gamma$, the admissible bandwidth is limited by $\ks/2$, since $\alpha^*_{\mathrm{abs}} \approx  \ks/2\Gamma$.

In \cref{fig:eta_vs_alpha}\textbf{(c)}, we compute the minimal temporal extent $1/\alpha^*_{\mathrm{abs}}\Gamma$ as a function of $\gens$ for different values of the intrinsic loss rate $\kappa_0$. Interestingly, the optimal operating bandwidth increases with intrinsic losses, but of course at the expense of protocol efficiency.
Moreover, the lower bound for this extent is given by $2/\Gamma$. This result has direct implications for the memory system. First, it limits its capacity to store multiple temporal modes. Indeed, at best, one would be able to store in the memory $\sim T_2\Gamma/2$ temporal pulses, and in the case of a hybrid system with a small ensemble coupling constant, only $\sim T_2 \ks/2$ pulses. In the case of a second-long coherence time, this metric appears very optimistic since the capacity would then largely exceed the number of spins in the ensemble, which goes beyond the hypothesis of storing few excitations compared to the number of spins laid out in \cref{sec:Physical model}. One should then consider another metric, which would account for the breakdown of the Holstein-Primakoff approximation.

This minimal temporal extent also imposes constraints for inserting the memory in a modular architecture. Let us consider the simplest architecture of one qubit able to emit on demand its quantum state in the input waveguide of the quantum memory, with a wavepacket envelope appropriate for perfect absorption into the memory. To be a faithful representation of its past state, the itinerant quantum state should have a temporal extent significantly smaller than the emitting qubit coherence times. However, if the corresponding bandwidth is too large, it may not be able to be absorbed with good efficiency into the memory, independently of its shape. Assuming a coherence time of $\SI{15}{\micro\second}$ for a qubit connected in a quantum state transfer type architecture \cite{Kurpiers2018}, it implies that the wavepacket should be emitted within the order of $\SI{1.5}{\micro\second}$. Assuming minimal intrinsic losses $\kappa_0/(2\pi)=\SI{20}{\kilo\hertz}$ and a spin linewidth $\Gamma/(2\pi)=\SI{300}{\kilo\hertz}$, the ensemble coupling constant should then be larger than $\gens/(2\pi)=\SI{120}{\kilo\hertz}$ to be able to absorb efficiently this wavepacket. This constraint appears very reasonable in light of the parameters we considered for promising spin systems.

\begin{figure}
    \centering
    \includegraphics[width=\linewidth]{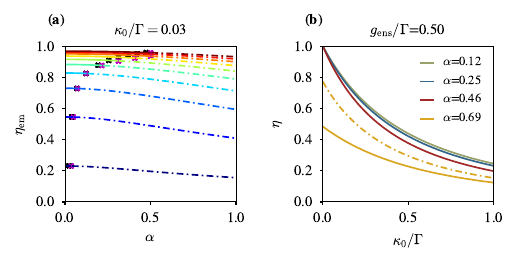}
    \caption{\textbf{(a)} Emission efficiency analysis as a function of the pulse speed $\alpha$ using the same convention as in \cref{fig:eta_vs_alpha}\textbf{(b)}. Dash-dotted line is the theoretical upper bound given by \cref{eq:UpperBound0}, that is saturated for $\alpha<\alpha^*_{\mathrm{em}}$ (magenta crosses), which closely matches $\alpha^*_{\mathrm{abs}}$ (black crosses).
    \textbf{(b)} Efficiency versus cavity internal losses for the four pulse speeds in \cref{fig:shapesComp}. For this coupling strength of $\gens=\Gamma/2$, we have  $\alpha_{\text{em}}^*\approx 0.5$ for all values of $\kappa_0$. Thus, the first three pulses remain below the critical speed and reach the upper bounds shown by green, blue and red solid curves. The fastest pulse exceeds this critical speed, so that the upper bound (dash-dotted) can not be reached. Taking the suboptimal choice of $\kappa_{\text{e}}(t)=\kappa_{\text{a}}(T_{\text{E}}-t)$ yields the solid yellow curve.}
    \label{fig:efficiencies}
\end{figure}

The above analysis only concerns the optimal absorption efficiency. Concerning the optimal emission efficiency,~\eqref{eq:UpperBound0} provides an upper bound. This upper bound is plotted in Fig.~\ref{fig:efficiencies}\textbf{(a)}. This upper bound is attained for slow enough pulses such that the cavity bandwidth modulation $\kappa_{\textrm{e}}^*$ given by~\eqref{eq:kappa2star} satisfies the positivity constraint. In Fig.~\ref{fig:efficiencies}\textbf{(a)}, this corresponds to $\alpha<\alpha_{\textrm{em}}^*$ denoted by magenta crosses. As can be seen in the plot, these critical values are very close to the critical values for the absorption process. Above these critical values, the upper bound is not necessarily saturated, and to find the actual optimal emission efficiency, one needs to solve a convex optimization problem with constraints as stated before.

We conclude by evaluating the impact of intrinsic losses on the total efficiency $\eta$ of the protocol. The numerical result is depicted in \cref{fig:efficiencies}\textbf{(b)}. The parameter $\gens/\Gamma$ is set to $0.5$ ($\alpha^*_{\textrm{em}}\sim \alpha^*_{\mathrm{abs}}\sim 0.5$ for all values of $\kappa_0$), and we plot the efficiency as a function of $\kappa_0/\Gamma$ for $\alpha=0.12,0.25,0.46$ and $\alpha=0.69>\alpha^*_{\mathrm{abs}},\alpha_{\textrm{em}}^*$. In this last case of pulse speed above the critical point, the optimal modulation for $\kappa_{\textrm{e}}$, given by \cref{eq:kappa2star}, violates the positivity constraint. Consequently, we adopt the suboptimal modulation $\kappa_{\textrm{e}}(t)=\kappa_{\textrm{a}}^*(T_E-t)$.
This choice is motivated by the fact that, for slow pulses, this time-reversed modulation is close to the optimal emission modulation (see \cref{fig:kappas}).
\cref{fig:efficiencies}\textbf{(b)} shows how much the
efficiency associated with this suboptimal modulation (yellow, solid) deviates from its upper bound (yellow, dash-dotted), which is given by the product of the optimal absorption efficiency and the upper bound for emission efficiency (see \cref{eq:UpperBound0}).

\section{Conclusion}
In this work, we have presented a quantum cascaded model for absorption and emission of an itinerant wavepacket into a spin ensemble embedded in a cavity. This model is derived in a mean-field framework and allows us to derive an upper bound for the memory catch and release efficiency depending on the hybrid system parameters. Spin coherence dynamics, for which we predict a limited impact, were excluded from the present derivation for conciseness, but could be straightforwardly included. We demonstrate that this efficiency bound is reached for adiabatic “slow" pulses and that there exists a critical bandwidth below which the memory performance remains close to this upper bound. Above this threshold, the efficiency decreases severely. We also derived the required modulation of the cavity coupling to the input waveguide to reach optimal absorption and emission of a given wavepacket.

Another assumption of our model that could be easily revisited is the hypothesis of perfect refocusing pulses. While this allowed us to express simply the feedback occurring between absorption and emission, a more precise description of the action of the pulses, depending on each spin frequency or coupling constant, could be taken into account, which would allow us to derive a finite-refocusing efficiency. 

We conclude from this analysis that achieving in practice a high-fidelity quantum memory will require a bandwidth-tunable cavity with low intrinsic losses. Concerning the spin ensemble and its coupling to the cavity, the required parameters are within the reach of current experimental conditions.

In future works, the results presented here could also be compared to numerical simulations discretizing the spin ensemble to study its dynamics~\cite{julsgaard} or to experimental realizations. Another theoretical development could be to use a mean-field framework extended to the second order to include the noise dynamics of the cavity and of the spin ensemble in the present analysis, and see whether this analysis would confirm the mean-field dynamics we have described here. Going further, outlining under which conditions the current mean-field dynamics would map to a full bosonic mode interaction dynamics, where the spin-cavity system could be described by a system density matrix, would be extremely helpful in the context of modeling a modular architecture. Indeed, this ability would allow for deriving transfer efficiency and fidelity in a very straightforward manner despite the high number of degrees of freedom in the overall system and its hybrid nature. 

\subsection*{Data availability}
The datasets generated during the current study are available in a \href{https://github.com/LindaGreggio98/Qmemory.git}{Github repository}.
\subsection*{Acknowledgments}
L. Greggio, A. Petrescu and M. Mirrahimi acknowledge funding from French ANR grant OCTAVES (ANR-21-CE47-0007).  T. Lorriaux acknowledges the support of France 2030 project QuanTEdu-France ANR-22-CMAS-0001, and A. Bienfait and T. Lorriaux acknowledge funding from the Plan France 2030 through the project ANR-22-PETQ-0003.

\subsection*{Author contributions}
L.G., M.M. and A.B. performed the theoretical analysis, with contributions of A.P. and T.L.. L.G. performed the numerical simulations. L.G. and A.B. drafted the manuscript, and all authors reviewed it.
\subsection*{Competing interests}
The authors declare no competing interests
\begin{appendix}
\section{Derivation of the EOMs}
\label{app:EOMs}

\begin{figure*}
    \centering
    \includegraphics[width=\linewidth]{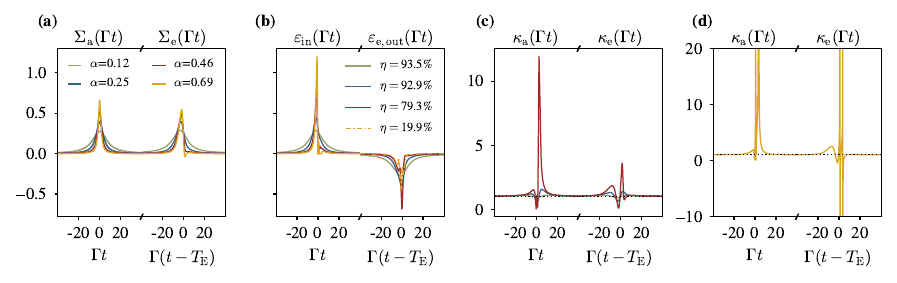}
    \caption{Same plots as in \cref{fig:shapesComp}, but for a Lorentzian pulse.
In \textbf{(c)}-\textbf{(d)} the printed efficiency for different values of the pulse speed is lower with respect to the relative one shown for a hyperbolic secant pulse (see \cref{fig:shapesComp}\textbf{(c)}-\textbf{(d)}), especially for fast pulses.
This demonstrates the better performance of a hyperbolic secant pulse compared to a Lorentzian pulse.}
    \label{fig:Lorentzian}
\end{figure*}

In this appendix, we derive the equations of motion for the intra-cavity field absorbed by the spins and then re-emitted at $T_{\text{E}}=2\tau_{\text{a}}+2\tau_{\text{e}}$.

We recall that during absorption ($t\leq \tau_{\text{a}}$), the dynamics of the intra-cavity field and of $\sigma^{j}_-$ of spin $j$ is dictated by the following EOMs
(see \cite{Afzelius_2013})
\begin{align}
   \frac{d}{d t} \varepsilon_{\text{a}}  = &-\frac{\kappa_{0}+\kappa_{\text{a}}(t)+i2\Delta_{\text{cs}}}{2} \varepsilon_{\text{a}} \nonumber \\
   &+i \sum_{j=1}^N g_j \sigma^{j,\text{a}}_-+\sqrt{\kappa_{\text{a}}(t)} \Ein(t), \label{eq:appAbsE} \\
\frac{d}{d t} \sigma^{j,\text{a}}_-  = &-i \Delta_j \sigma^{j,\text{a}}_-+ i g_j \varepsilon_{\text{a}}.  \label{eq:appAbsSigma}
\end{align}
We consider $\varepsilon_{\text{a}}(-\infty)=\sigma^{j,\text{a}}_-(-\infty)=0$ and we integrate \cref{eq:appAbsSigma}
\begin{equation}
\sigma^{j,\text{a}}_-(t)= i g_j \int_{-\infty}^{t}  ds \ e^{-i \Delta_j(t-s)} \varepsilon_{\text{a}}(s).
    \label{eq:spinsApp}
\end{equation}
In the EOM for the intra-cavity field (\ref{eq:appAbsE}), we replace the term $\sum_{j=1}^N g_j \sigma^{j}_-$ with its integrated version, using the spin frequency and coupling distributions
\begin{equation}
    \sum_{j=1}^N g_j \sigma^{j}_- \rightarrow \int_g \int_\Delta  g \sigma_-^{\Delta,g} p(g) n(\Delta)  dg d\Delta,
    \label{eq:sumToInt}
\end{equation}
and we plug in the expression for $\sigma_-^{\Delta,g}$ given in \cref{eq:spinsApp}, to obtain
\begin{equation*}
    - \gens^{2}   \int_{-\infty}^{t} d s\varepsilon_{\text{a}}(s)  \int d \Delta n(\Delta) e^{-i \Delta(t-s)},
\end{equation*}
with $\gens^2=\int p(g)g^2 dg$, and where we recognize the Fourier transform of the frequency distribution $n(\Delta)$, \ie $n(t-s)=e^{-\Gamma|t-s|/2}$. We can thus rewrite the above expression as $-\gens^2 \int_{-\infty}^{t} d s e^{-\Gamma|t-s| / 2} \varepsilon_{\text{a}}(s),$ that we equalize to $-\gens \Sigma_{\text{a}},$ where $\Sigma_{\text{a}}(t)= \gens \int_{-\infty}^{t} ds  e^{-\frac{\Gamma}{2}(t-s)} \varepsilon_{\text{a}}(s),$ with $  \Sigma_{\text{a}}(-\infty)=0,$ representing the field stored in the cavity.

Consequently, the spin-cavity EOMs during absorption rewrite as
\begin{equation}
\begin{split}
\frac{d}{d t} \varepsilon_{\text{a}}  =&-\frac{\kappa_{0}+\kappa_{\text{a}}(t)+i2\Delta_{\text{cs}}}{2} \varepsilon_{\text{a}}- \gens \Sigma_{\text{a}} \\
& +\sqrt{\kappa_{\text{a}}(t)} \Ein, \\
\frac{d}{d t} \Sigma_{\text{a}}  = & \gens \varepsilon_{\text{a}} -\frac{\Gamma}{2} \Sigma_{\text{a}}. 
\end{split} 
\label{eq:Abs2app}
\end{equation}
Now, we present the EOM for $\sigma^{j}_-$ of spin $j$ in between the two pulses ($ \tau_{\text{a}} < t < T_{\text{E}}=2\tau_{\text{a}}+\tau_{\text{e}}$), when the cavity is detuned
\begin{equation}
\begin{split}
\frac{d}{d t} \sigma^{j,\text{a/e}}_- & =+i \Delta_j \sigma^{j,\text{a/e}}_-
\end{split}
\label{eq:eqs2app}
\end{equation}
where the change of sign reflects the flipping caused by the first refocusing pulse.
Using the convention $\sigma^{j,\text{a/e}}_-(\tau_{\text{a}}^+)=\sigma^{j,\text{a}}_-(\tau_{\text{a}}^-)$ together with \cref{eq:spinsApp}, we obtain
\begin{align}
\sigma^{j,\text{a/e}}_-(t) &= \sigma^{j,\text{a/e}}_-(\tau_{\text{a}}) e^{i \Delta_j(t-\tau_{\text{a}})} \nonumber \\
&= i g_j \int_{-\infty}^{\tau_{\text{a}}}  ds \ e^{i \Delta_j(s+t-2\tau_{\text{a}})} \varepsilon_{\text{a}}(s). 
\label{eq:spinsInbet}
\end{align}
The EOMs during emission ($t \geq T_{\text{E}}=2\tau_a+\tau_e$) write as the ones during absorption (\ref{eq:appAbsSigma}), but without the driving term in the intra-cavity field EOM
\begin{align}
  \frac{d}{d t} \varepsilon_{\text{e}} & =-\frac{\kappa_{0}+\kappa_{\text{e}}(t)+i 2\Delta_{\text{cs}}}{2} \varepsilon_{\text{e}}+i \sum_{j=1}^N g_j \sigma_-^{j,\text{e}},  \label{eq:appEmE}\\
\frac{d}{d t} \sigma_-^{j,\text{e}} & =-i \Delta_j \sigma_-^{j,\text{e}}+ i g_j \varepsilon_{\text{e}}.   \label{eq:appEmSigma}
\end{align}
We integrate \cref{eq:appEmSigma} using the convention $\sigma_-^{j,\text{e}}(T_{\text{E}}^+)=\sigma^{j,\text{a/e}}_-(T_{\text{E}}^-)$ and we obtain
\begin{align}
 \sigma_-^{j,\text{e}}(t) =& \sigma_-^{j,\text{e}}(T_{\text{E}}) e^{i \Delta_j(T_{\text{E}}-t)} +i g_j \int_{T_{\text{E}}}^{t}  ds e^{i \Delta_j(s-t)} \varepsilon_{\text{e}}(s) \nonumber\\
 =& i g_j \int_{-\infty}^{\tau_{\text{a}}}  ds e^{i \Delta_j(s+T_{\text{E}}-t)} \varepsilon_{\text{a}}(s) \nonumber \\
 &+i g_j \int_{T_{\text{E}}}^{t}  ds e^{i \Delta_j(s-t)} \varepsilon_{\text{e}}(s), \label{eq:spinsEm}
\end{align}
where we have used \cref{eq:spinsInbet}. As done for absorption,  we replace the term $\sum_{j=1}^N g_j \sigma^{j}_-$ with its integrated version (see \cref{eq:sumToInt}), we substitute $\sigma_-^{\Delta,g}$ with its expression
(\ref{eq:spinsEm}), we recognize the Fourier transform of $n$, and we get
\begin{equation}
\begin{split}
  - \gens^2 &\left(  \int_{-\infty}^{\tau_{\text{a}}}  ds e^{- \frac{\Gamma}{2}|t-T_{\text{E}}-s|} \varepsilon_{\text{a}}(s)  \right.\\
  &\left. + \int_{T_{\text{E}}}^{t}  ds e^{- \frac{\Gamma}{2}(t-s)} \varepsilon_{\text{e}}(s) \right).       
\end{split}
 \label{eq:appEmterm}
\end{equation}
The second term can be expressed as $-\gens \tilde{\Sigma}_{\text{e}}$, which introduces a new variable $\tilde{\Sigma}_{\text{e}}(t)=\gens \int_{T_{\text{E}}}^{t}  ds e^{- \frac{\Gamma}{2}(t-s)} \varepsilon_{\text{e}}(s)$, with $\tilde{\Sigma}_{\text{e}}(T_{\text{E}})=0.$ To deal with the first integral we define the variable $\tilde \Sigma_{\text{a}}(t)=\gens \int_{-\tau_{\text{a}}}^{t} ds e^{- \frac{\Gamma}{2}(t-s)} \varepsilon_{\text{a}}(-s),$
with $\tilde \Sigma_{\text{a}}(-\tau_{\text{a}})=0;$  representing a filtered version of the field stored in the spins during absorption. Using these new variables, together with the previously defined $\Sigma_{\text{a}}$, we re-express \cref{eq:appEmterm} as $-\gens \Sigma_{\text{e}}$, where the field stored in the spin ensemble during emission is defined as
\begin{equation*}
    \Sigma_{\text{e}}(t)=\begin{cases}
        \Sigma_{\text{a}}(t-T_{\text{E}}) +\tilde \Sigma_{\text{a}}(T_{\text{E}}-t) + \tilde \Sigma_{\text{e}}(t) \ &\text{if } t\leq T_{\text{E}}+\tau_{\text{a}} \\
        \Sigma_{\text{a}}(\tau_{\text{a}})e^{-\frac{\Gamma}{2}(t-T_{\text{E}}-\tau_{\text{a}})}+\tilde \Sigma_{\text{e}}(t) \ &\text{if } t> T_{\text{E}}+\tau_{\text{a}}
    \end{cases}
\end{equation*}
We can now present the EOMs for the field in the cavity and the one stored in the spins during emission
\begin{equation}
\begin{split}
\frac{d}{d t} \varepsilon_{\text{e}}  &= -\frac{\kappa_{0}+\kappa_{\text{e}}(t)+i2 \Delta_{\text{cs}}}{2} \varepsilon_{\text{e}}  - \gens \Sigma_{\text{e}},\\
\frac{d}{d t}  \Sigma_{\text{e}} &= \begin{cases}
        \gens \varepsilon_{\text{e}}-\frac{\Gamma}{2} \Sigma_{\text{e}}+\Gamma \tilde \Sigma_{\text{a}}(T_{\text{E}}-t) \ &\text{if } t\leq T_{\text{E}}+\tau_{\text{a}} \\
        \gens \varepsilon_{\text{e}}-\frac{\Gamma}{2} \Sigma_{\text{e}} \ &\text{if } t> T_{\text{E}}+\tau_{\text{a}} 
    \end{cases}\\
    \frac{d}{d t} \tilde{\Sigma}_{\text{a}} &=-\frac{\Gamma}{2}\tilde{\Sigma}_{\text{a}}+\gens \varepsilon_{\text{a}}(-t),
\end{split}
\label{eq:EmApp}
\end{equation}
with $\tilde \Sigma_{\text{a}}(T_{\text{E}}-t)$ representing a feedback term from absorption, appearing when $t\leq T_{\text{E}}+\tau_{\text{a}}$. 

Concerning $\Sigma_{\text{e}}(t)$, continuity and differentiability are not lost in $t=T_{\text{E}}+\tau_{\text{a}}$.
The initial conditions are $\varepsilon_{\text{e}}(T_{\text{E}})=\varepsilon_{\text{a}}(\tau_{\text{a}})e^{-\frac{\kappa_0}{2}(\tau_{\text{a}}+\tau_{\text{e}})}, \
      \Sigma_{\text{e}}(T_{\text{E}})= \Sigma_{\text{a}}(-\tau_{\text{e}})+\tilde{\Sigma}_{\text{a}}(\tau_{\text{e}}).$

In our work, we neglect the cut-off of the feedback term since it does not affect the storage time interval, and we consider $\Sigma_{\text{e}}$ undergoing the following EOMs
  \begin{equation}
     \dot{\Sigma}_{\text{e}}=\gens \varepsilon_{\text{e}}-\frac{\Gamma}{2}\Sigma_{\text{e}}+\sqrt{\Gamma} \Sigma_{\text{e,in}},
     \label{eq:appSigma2}
 \end{equation}
 with $\Sigma_{\text{e,in}}=\sqrt{\Gamma}\tilde \Sigma_{\text{a}}(T_{\text{E}}-t)$ being the feedback term from absorption.

 Finally, we consider $\tau_{\text{a}}$ large enough such that the entire field has left the cavity at the moment of the first refocusing pulse. This allows us to consider the intra-cavity field  $\varepsilon_{\text{a}}$ to be zero after $\tau_{\text{a}}$, and the intra-cavity field $\varepsilon_{\text{e}}$ to be zero before $T_{\text{E}}.$ The same consideration holds for the field stored in the spins since we can consider that before the refocusing pulses the information has been totally washed out by inhomogeneous broadening, namely $\Sigma_{\text{a}}$ and $\Sigma_{\text{e}}$ are zero after $\tau_{\text{a}}$ and before $T_{\text{E}}$ respectively. Under this assumption, we can consider $\varepsilon(\pm \infty)=\Sigma(\pm \infty)=0.$

\section{Transmission and reflection coefficients}
\label{app:coeffs}
In this appendix, we aim to derive the expression for the transmission and reflection coefficients in the frequency domain, in the case of constant coupling $\kappa$ and detuning $\Delta_{\text{cs}}.$

We recall the definition of the transmission and reflection coefficients associated with absorption and emission
\begin{equation}
        \begin{split}
      t_{\text{a}}=\Sigma_{\text{a,out}}/\Ein  , \ &t_{\text{e}}=\varepsilon_{\text{e,out}}/\Sigma_{\text{e,in}},\\
      r_{\text{a}}=\varepsilon_{\text{a,out}}/\Ein, \ &r_{\text{e}}=\Sigma_{\text{e,out}}/\Sigma_{\text{e,in}},
    \end{split}
    \label{eq:trcoeff}
\end{equation}
with $t_{\text{a}}$ being the transmission coefficient from the driveline
to the spins, $t_{\text{e}}$ being the one from the spins back to the driveline, $r_{\text{a}}$ being the reflection coefficient back to the driveline and $r_{\text{e}}$ being the one back to the spins. 

We rewrite the absorption EOMs (\ref{eq:Abs2app}) in the frequency domain, considering a non-zero detuning $\Delta_{\text{cs}}$ and a constant $\kappa_{\text{a}}$
\begin{equation}
\begin{split}
i \delta \varepsilon_{\text{a}}[\delta]  = &-i 
\Delta_{\text{cs}} \varepsilon_{\text{a}}[\delta] -\frac{\kappa_{0}+\kappa_{\text{a}}}{2} \varepsilon_{\text{a}}[\delta]- \gens \Sigma_{\text{a}}[\delta] \\
&+\sqrt{\kappa_{\text{a}}} \Ein[\delta], \\
i \delta \Sigma_{\text{a}}[\delta]  = & \gens \varepsilon_{\text{a}}[\delta] -\frac{\Gamma}{2} \Sigma_{\text{a}}[\delta]. 
\end{split} 
\label{eq:Absfreq}
\end{equation}
Rearranging the terms we get $\Sigma_{\text{a}}[\delta]$ and $\varepsilon_{\text{a}}[\delta]$ as functions of $\Ein[\delta]$
\begin{equation}
\begin{split}
       \Sigma_{\text{a}}[\delta]&=\frac{\frac{4 \gens}{\Gamma}\sqrt{\kappa_{\text{a}}}}{d[\delta,\Delta_{\text{cs}},\kappa_{\text{a}}]}\Ein[\delta], \\
       \varepsilon_{\text{a}}[\delta] &= \frac{2\sqrt{\kappa_{\text{a}}}\left(1+i \frac{2\delta}{\Gamma}\right)}{d[\delta,\Delta_{\text{cs}},\kappa_{\text{a}}]}\Ein[\delta],
\end{split}
\label{eq:Sigma1Ein}
\end{equation}
with
\begin{equation}
d[\delta,\Delta_{\text{cs}},\kappa]=\kappa_0+\kappa+\tilde{\kappa}_{\text{s}} + 2 i (\tilde{\delta}+\Delta_{\text{cs}})
\end{equation}
where $\ks=4 \gens^2/\Gamma$, $\tilde{\kappa}_{\text{s}}/\ks = 1-\frac{\delta(\delta+\Delta_{\text{cs}})}{\gens^2}$ and $\tilde{\delta}/\delta = 1+ (\kappa_0+\kappa)/\Gamma$.
We consider the absorption transmission and reflection coefficients defined in \cref{eq:trcoeff} in the frequency domain, we substitute $\Sigma_{\text{a,out}}$ with $\sqrt{\Gamma}\Sigma_{\text{a}}$ and $\varepsilon_{\text{a,out}}$ with $\sqrt{\kappa_{\text{a}}(t)}\varepsilon_{\text{a}}-\Ein$,
and we plug in \cref{eq:Sigma1Ein} to obtain
\begin{equation}
    \begin{split}
      t_{\text{a}}[\delta]&=\frac{2\sqrt{\ks \kappa_{\text{a}}}}{d[\delta,\Delta_{\text{cs}},\kappa_{\text{a}}]}, \\
      r_{\text{a}}[\delta] &=\frac{2 \kappa_{\text{a}} \left(1+i \frac{2\delta}{\Gamma}\right)}{d[\delta,\Delta_{\text{cs}},\kappa_{\text{a}}]}-1.
    \end{split}
    \label{eq:appcoeffAbs}
\end{equation}

We pass to emission and we rewrite the relative EOMs (first eq. in (\ref{eq:EmApp}) and \cref{eq:appSigma2}) in the frequency domain, considering the detuning $\Delta_{\text{cs}}$ and a constant $\kappa_{\text{e}}$
\begin{equation}
    \begin{split}
     i \delta \varepsilon_{\text{e}}[\delta]  &=-i \Delta_{\text{cs}} \varepsilon_{\text{e}}[\delta]  -\frac{\kappa_{0}+\kappa_{\text{e}}}{2} \varepsilon_{\text{e}}[\delta]  - \gens \Sigma_{\text{e}}[\delta],\\   
     i \delta  \Sigma_{\text{e}}[\delta] &=  \gens \varepsilon_{\text{e}}[\delta]-\frac{\Gamma}{2} \Sigma_{\text{e}}[\delta]+\sqrt{\Gamma} \Sigma_{\text{e,in}}[\delta].
    \end{split}
    \label{eq:Emfreq}
\end{equation}
Rearranging the terms we get $\Sigma_{\text{e}}[\delta]$ and $\varepsilon_{\text{e}}[\delta]$ as functions of the driving field $\Sigma_{\text{e,in}}[\delta]$
\begin{equation}
\begin{split}
     \Sigma_{\text{e}}[\delta] &= \frac{\frac{2}{\sqrt{\Gamma}}(2 i (\delta+\Delta_{\text{cs}})+ \kappa_{0}+\kappa_{\text{e}})}{d[\delta,\Delta_{\text{cs}},\kappa_{\text{e}}]} \Sigma_{\text{e,in}}[\delta],\\
     \varepsilon_{\text{e}}[\delta] &= -\frac{2 \sqrt{\ks}}{d[\delta,\Delta_{\text{cs}},\kappa_{\text{e}}]}\Sigma_{\text{e,in}}[\delta].
     \label{eq:Sigma2Ein}
\end{split}
\end{equation}
We consider the emission transmission and reflection coefficients defined in \cref{eq:trcoeff} in the frequency domain, substituting $\varepsilon_{\text{e,out}}$ with $\sqrt{\kappa_{\text{e}}}\varepsilon_{\text{e}}$ and $\Sigma_{\text{e,out}}$ with $\sqrt{\Gamma}\Sigma_{\text{e}}-\Sigma_{\text{e,in}}$,
and we plug in \cref{eq:Sigma2Ein} to obtain
\begin{equation}
    \begin{split}
      t_{\text{e}}[\delta]&= -\frac{2 \sqrt{\kappa_{\text{e}} \ks}}{d[\delta,\Delta_{\text{cs}},\kappa_{\text{e}}]}, \\
      r_{\text{e}}[\delta]&= \frac{2(2 i (\delta+\Delta_{\text{cs}})+ \kappa_{0}+\kappa_{\text{e}})}{d[\delta,\Delta_{\text{cs}},\kappa_{\text{e}}]}-1.
    \end{split}
    \label{eq:appcoeffEm}
\end{equation}
Note that the transmission coefficient during emission represents a mirrored version of the one during absorption, \ie $t_{\text{e}}[\delta]=-t_{\text{a}}[\delta],$ with relative coupling constant $\kappa_{\text{e}}$ and $\kappa_{\text{a}}.$

\section{Input-output energies}
\label{app:inOutEnergies}

In this appendix, we provide the derivation of the inequality~\eqref{eq:disEin} leading to the upper bound for the absorption process, and the identity~\eqref{eq:output_en} required for treating the optimization of the emission process. 

We recall that the input energy evaluated at the optimal $\bar{\kappa}_{\text{a}}^*$ reads
\begin{equation}
    \min \int \Eein(\tau)^2 d\tau = \int \left| \frac{B}{A}\right| \left(A+ B \left| \frac{A}{B}\right|\right)^2 d\tau,
\end{equation}
where $A({\bar{\kappa}_{\text{a}}})$ and $B({\bar{\kappa}_{\text{a}}})$ are defined in \cref{eq:ABdefinition}.
The integrand $\left| \frac{B}{A}\right| \left(A+ B \left| \frac{A}{B}\right|\right)^2$ can be rewritten as $2|AB|+2AB$, which is lower bounded by $4AB$. Consequently, a lower bound for the input energy is given by
\begin{equation*}
    \begin{split}
       4& \int d\tau A B  =  \\
       =& \frac{1}{4\bargens^2}  \int d\tau [ 2(4\ddot{\mathcal{S}}_{\text{a}} \dot{\mathcal{S}}_{\text{a}} + 2(1 + \bar{\kappa}_0 )\dot{\mathcal{S}}_{\text{a}}^2  +( \bar{\kappa}_0+\bar{\kappa}_{\text{s}}) \mathcal{S}_{\text{a}} \dot{\mathcal{S}}_{\text{a}} ) \\
       &  +4 \ddot{\mathcal{S}}_{\text{a}} \mathcal{S}_{\text{a}} + 2(1 + \bar{\kappa}_0 )\dot{\mathcal{S}}_{\text{a}} \mathcal{S}_{\text{a}} +( \bar{\kappa}_0+\bar{\kappa}_{\text{s}}) \mathcal{S}_{\text{a}}^2  ] \\
       = &\frac{1}{\bar{\kappa}_{\text{s}}}   \int d\tau [4\bar{\kappa}_0 \dot{\mathcal{S}}_{\text{a}}^2 +( \bar{\kappa}_0+\bar{\kappa}_{\text{s}}) \mathcal{S}_{\text{a}}^2],
    \end{split}
\end{equation*}
where we have used the fact that $ \mathcal{S}_{\text{a}}(\pm \infty)=\dot{\mathcal{S}}_{\text{a}}(\pm \infty)=0.$ This proves the inequality~\eqref{eq:disEin}. 

Now, for the emission process, we recall the EOMs
\begin{equation}
\begin{split}
\frac{d}{d \tau} \mathcal{E}_{\text{e}}  &= -\frac{\bar{\kappa}_{0}+\bar{\kappa}_{\text{e}}(\tau)}{2} \mathcal{E}_{\text{e}}  - \mathfrak{g} \mathcal{S}_{\text{e}},\\
\frac{d}{d \tau}  \mathcal{S}_{\text{e}} &=-\frac{1}{2}\mathcal{S}_{\text{e}}+ \bargens \mathcal{E}_{\text{e}}+ \mathcal{S}_{\text{in}}(\bar T_E-\tau).\\
\end{split}
\label{eq:simplEmapp}
\end{equation}
We multiply the first equation by $\mathcal{E}_{\text{e}}$, and the second equation by $\mathcal{S}_{\text{e}}$, then integrate both of them over time to obtain 
 \begin{equation}
    \begin{split}
     \int d\tau \mathcal{E}_{\text{e}} \dot{\mathcal{E}_{\text{e}}}  &= \int d\tau \left ( -\frac{\bar{\kappa}_{0}+\bar{\kappa}_{\text{e}}(\tau)}{2} \mathcal{E}_{\text{e}}^2 - \bargens \mathcal{S}_{\text{e}} \mathcal{E}_{\text{e}} \right),\\   
          \int d\tau \mathcal{S}_{\text{e}} \dot{\mathcal{S}}_{\text{e}} &= \int d\tau \left(  \bargens \mathcal{E}_{\text{e}} \mathcal{S}_{\text{e}}-\frac{1}{2} \mathcal{S}_{\text{e}}^2+ \mathcal{S}_{\text{e}} \mathcal{S}_{\text{in}} \right).
    \end{split}
    \label{eq:emission2int}
\end{equation}
We recall that $\mathcal{E}_{\text{e}}(\pm \infty)=\mathcal{S}_{\text{e}}(\pm \infty)=0,$ we sum the two equations in (\ref{eq:emission2int}) and we get
\begin{equation}
    0=\int d\tau \left(-\frac{\bar{\kappa}_0+\bar{\kappa}_{\text{e}}(\tau)}{2}\mathcal{E}_{\text{e}}^2-\frac{1}{2}\mathcal{S}_{\text{e}}^2+  \mathcal{S}_{\text{e}} \mathcal{S}_{\text{in}} \right),
\end{equation}
from which we derive the output energy as
\begin{equation}
    \int d\tau \bar{\kappa}_{\text{e}}(\tau) \mathcal{E}_{\text{e}}^2 = \int d\tau \left(-\bar{\kappa}_0 \mathcal{E}_{\text{e}}^2-  \mathcal{S}_{\text{e}}^2+2  \mathcal{S}_{\text{e}} \mathcal{S}_{\text{in}}  \right).
    \label{eq:outEn}
\end{equation}
We rewrite the second equation in (\ref{eq:simplEmapp}) as $\mathcal{E}_{\text{e}} = \frac{1}{\bargens}  \left( \dot{\mathcal{S}}_{\text{e}}+\frac{1}{2}\mathcal{S}_{\text{e}}- \mathcal{S}_{\text{in}}\right)$,
we insert it in \cref{eq:outEn} and we obtain
\begin{align}
    \int d\tau \bar{\kappa}_{\text{e}}(\tau) \mathcal{E}_{\text{e}}^2 =\int d\tau \left(-\frac{\bar{\kappa}_0}{\bargens^2} \left( \dot{\mathcal{S}}_{\text{e}}+\frac{1}{2}\mathcal{S}_{\text{e}} \nonumber  - \mathcal{S}_{\text{in}}\right)^2 \right. \\  \left. -  \mathcal{S}_{\text{e}}^2+2  \mathcal{S}_{\text{e}} \mathcal{S}_{\text{in}}  \right).
    \label{eq:outEn2}
\end{align}
To pass to the frequency domain through Parseval's theorem, we get rid of the cross product $\mathcal{S}_{\text{e}} \mathcal{S}_{\text{in}}$ by rewriting the term $- \mathcal{S}_{\text{e}}^2+2  \mathcal{S}_{\text{e}} \mathcal{S}_{\text{in}}$ as $- (\mathcal{S}_{\text{e}}-\mathcal{S}_{\text{in}})^2+ \mathcal{S}_{\text{in}}^2.$ 
Consequently, we obtain the identity~\eqref{eq:output_en}.


\section{Further numerical results}
\label{app:numerics}
\begin{figure}
    \centering
    \includegraphics[width=\linewidth]{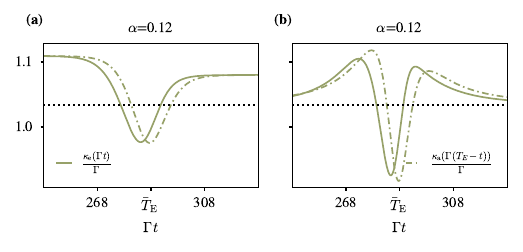}
    \caption{Same conventions as in of \cref{fig:shapesComp}. Optimal shape for $\kappa_{\text{e}}(t)$ (solid line, as defined in \cref{eq:kappa2star}) compared to the time-reversed absorption profile $\kappa_{\text{a}}(T_{\text{E}} - t)$ (dashed-dotted line), for a slow hyperbolic secant pulse (\textbf{a}) and a slow Lorentzian pulse (\textbf{b}), both not far from the the “slow-pulse” limit (black-dotted line). When $\alpha < \alpha^*_{\text{em}}$, $\kappa_{\text{e}}(t)$ closely resembles a mirrored version of the time-reversed $\kappa_{\text{a}}$, supporting the choice of using a mirrored coupling profile for emission in the fast-pulse regime ($\alpha \geq \alpha^*_{\text{em}}$).}
    \label{fig:kappas}
\end{figure}
In this appendix, we present the simulation results for a Lorentzian pulse, showing its worse performance compared to the chosen hyperbolic secant pulse in terms of efficiency. Moreover, we show that, for $\alpha < \alpha_{\text{em}}^*$, the optimal modulation for the coupling during emission is approximately a mirrored version of the one derived for the absorption step, justifying why we are taking this approximation above the critical speed, when the positivity constraint for the optimization problem should be active.

The Lorentzian pulse, normalized to unity, writes $\Sigma_{\text{a}}(t)=\frac{\sqrt{2\alpha/\pi}}{1+(\alpha \Gamma t)^2}$. The numerics are implemented the same way as for the hyperbolic secant pulse (see \cref{subsec:hsPulse}). The results are presented in \cref{fig:Lorentzian}, showing a lower efficiency compared to the one obtained by the corresponding hyperbolic secant pulse.

The optimal $\kappa_{\text{e}}$, given by \cref{eq:kappa2star}, for both types of pulses is presented in \cref{fig:kappas}, together with the time-reversed version of the optimal $\kappa_{\text{a}}$, given by \cref{eq:kappa1star}. The first one represents a slightly translated version of the second one, which allows us to perform the approximation.  

\section{Symbols and notations}

\begin{table}[h!]
\centering
\begin{tabular}{cl}
\toprule
&\textbf{General Notation and Constants}\\
\midrule
$\omega_s$ & Mean spin Larmor frequency \\
$\omega_c$ & Cavity frequency \\
$\omega_j$ & Larmor frequency of spin $j$ \\
$\Delta_j$ & Detuning of spin $j$ from $\omega_s$: $\Delta_j = \omega_j - \omega_s$ \\
$\Delta_{\text{cs}}$ & Detuning between cavity and spin central \\
&frequency: $\Delta_{\text{cs}} = \omega_c - \omega_s$ \\
$\Gamma$ & Inhomogeneous linewidth of the spin ensemble \\
$T_2$ & Coherence time of the spin ensemble \\
$T_E$ & Echo time: $T_E = 2\tau_a + 2\tau_e$ \\
$\tau_a$ & Duration of the absorption step \\
$\tau_e$ & Duration of the emission step \\
\midrule
\\
& \textbf{Operators}\\
\midrule
$\hat{\varepsilon}$ & Annihilation operator for the cavity field \\
$\hat{\varepsilon}^\dagger$ & Creation operator for the cavity field \\
$\hat{\sigma}^j_-$ & Lowering operator for spin $j$ \\
$\hat{\sigma}^j_+$ & Raising operator for spin $j$ \\
$\hat{\sigma}^j_z$ & Pauli-$z$ operator for spin $j$ \\
\midrule
\\
& \textbf{Fields and Couplings}\\
$\Ein(t) [\mathcal{E}_{\mathrm{in}}]$ & Input field envelope \\
$\Eout(t) [\mathcal{E}_{\mathrm{out}}]$ & Output field envelope \\
$\varepsilon(t) [\mathcal{E}]$& Intra-cavity field \\
$\Sigma(t) [\mathcal{S}]$ & Field in the spin ensemble \\
$g_j$ & Coupling strength between cavity and spin $j$ \\
$\gens [\mathfrak{g}]$ & Collective coupling strength: $\gens^2 = \int p(g)g^2 \, dg$ \\
$\kappa_0 [\bar{\kappa}_0]$ & Intrinsic loss rate of the cavity \\
$\kappa(t) [\bar{\kappa}]$& Coupling rate of the cavity to the input line \\
$\kappa_{\text{a}}(t) [\bar{\kappa}_{\text{a}}]$ & Cavity coupling rate during absorption \\
$\kappa_{\text{e}}(t) [\bar{\kappa}_{\text{e}}]$ & Cavity coupling rate during emission \\
$\ks  [\bar{\kappa}_{\text{s}}]$ & Spin-induced loss rate on the cavity: $\ks = 4\gens^2/\Gamma$ \\
$\Sigma_{\text{a}} [\mathcal{S}_{\text{a}}]$ & Spin field during absorption \\
$\Sigma_{\text{e}} [\mathcal{S}_{\text{e}}]$ & Spin field during emission \\
\midrule
\\
&\textbf{Efficiency and Transmission Coefficients}\\
$\eta$ & Overall efficiency of the protocol $\eta_a \times \eta_H \times \eta_e$ \\
$\eta_a$ & Absorption efficiency \\
$\eta_e$ & Emission efficiency \\
$\eta_H$ & Efficiency of the refocusing pulses \\
$t_a$ & Transmission coef. from driveline to spins\\
$t_e$ & Transmission coef. from spins back to driveline\\
$r_a$ & Reflection coef. back to the driveline \\
$r_e$ & Reflection coef. back to the spins \\
\midrule
\\
& \textbf{Optimization and Mathematical Functions}\\
$H[\delta]$ & Filter function between $\Sigma_{\text{a,out}}$ and $\Sigma_{\text{e,in}}$ \\
$\mathcal{F}$ & Fourier transform \\
$\mathcal{F}^{-1}$ & Inverse Fourier transform \\
$\bar{\kappa}_a^*$ & Optimal cavity coupling rate during absorption \\
$\bar{\kappa}_e^*$ & Optimal cavity coupling rate during emission \\
\midrule
\\
& \textbf{Physical Parameters for Simulations}\\
$\alpha$ & Bandwidth of the input pulse: $\alpha \Gamma$ \\
$\alpha_{\text{abs}}^*$ & Critical bandwidth for absorption \\
$\alpha_{\text{em}}^*$ & Critical bandwidth for emission \\
$\kappa_{\text{max}}$ & Maximum coupling rate \\
$\kappa_{\text{min}}$ & Minimum coupling rate \\
\bottomrule
\end{tabular}
\caption{List of symbols and their definitions. Symbol $X$ in $[X(\tau)]$ indicates a rescaled quantity by the change of time variable $\tau=\Gamma t$.}
\end{table}

\end{appendix}

\bibliographystyle{apsrev4-1}
\bibliography{bibliography}
\end{document}